\documentclass[12pt]{article}
\pdfoutput=1
\usepackage{amssymb,graphicx}
\usepackage{epstopdf}
\usepackage{amsmath,amsfonts}
\usepackage{epsfig}
\usepackage{graphicx,graphics}

\begin{document}

\sloppy

\title{\bf Gravitational focusing of\\ Imperfect Dark Matter}
\author{Eugeny Babichev$^a$\footnote{{\bf e-mails}:
   eugeny.babichev@th.u-psud.fr, sabir.ramazanov@gssi.infn.it}\ ~and Sabir Ramazanov$^{b}$\\
 \small{$^a$\em Laboratoire de Physique Th\'eorique, CNRS,} \\ 
 \small{\em  Univ. Paris-Sud, Universit\'e Paris-Saclay, 91405 Orsay, France}\\
  \small{$^b$\em Gran Sasso Science Institute (INFN), Viale Francesco Crispi 7, I-67100 L'Aquila, Italy}\\
 }

{\let\newpage\relax\maketitle}

\begin{abstract}
Motivated by the projectable Horava--Lifshitz model/mimetic matter scenario, we consider a particular modification of 
standard gravity, which manifests as an imperfect low pressure fluid. 
While practically indistinguishable from a collection of 
non-relativistic weakly interacting particles on cosmological scales, it leaves drastically different 
signatures in the Solar system. The main effect stems from gravitational focusing of the flow of {\it Imperfect Dark Matter} 
passing near the Sun. This entails strong amplification of Imperfect Dark Matter energy density compared to  its average value in the surrounding halo. 
The enhancement is many orders of magnitude larger than in the case of Cold Dark Matter, provoking deviations of the metric in the second order in the Newtonian potential. Effects of gravitational focusing are prominent 
enough to substantially affect the planetary dynamics. Using the existing bound on the PPN parameter $\beta_{PPN}$, we deduce a stringent constraint on the unique constant of the model.

\end{abstract}

\section{Introduction and Summary}

Despite successes of General Relativity (GR)~\cite{Willbook, Will:2014kxa}, it appears to be 
incomplete in both high and low energy limits. First, GR is not perturbatively renormalizable and, consequently, 
loses its predictive power at distances of the order of the Planckian size, $l_{Pl} \sim 10^{-33}~\mbox{cm}$. 
To retain predictivity at those and smaller scales, 
one should replace GR by some ultraviolet complete theory {\it a la} superstrings. The 'infrared' problem stems from the existence of Dark Energy,---we yet do not 
know the physics behind the small $\Lambda$-constant. 
Another low energy phenomenon, which is less commonly viewed 
as challenge for GR, is Dark Matter (DM). While it could be relatively simply explained in some extensions of the Standard Model, 
only gravitational manifestations of DM have been identified so far. Therefore, it may be well a product of  gravity modification. We will 
entertain this possibility in the present paper.

We will be interested in the model  of gravity described by the following action~\cite{Chamseddine:2014vna, Capela:2014xta, Mirzagholi:2014ifa}, 
\begin{equation}
\label{action}
S= -\frac{1}{16\pi G} \int d^4 x \sqrt{-g} R+\int d^4 x \sqrt{-g} \left[\frac{\Sigma}{2} \left( g^{\mu \nu} \partial_{\mu} \varphi \partial_{\nu} \varphi-1 \right)+\frac{\chi}{2} (\square \varphi )^2 \right] \; ,
\end{equation}
(hereafter, we assume the mostly negative signature of the metric). The first term on the r.h.s. here describes the Einstein--Hilbert action; $G \equiv \frac{1}{M^2_{Pl}}$ is the gravitational constant, and $M_{Pl}$ is the Planck mass. The other two stand for what one calls {Imperfect Dark Matter} (IDM). In the 
limit of vanishing dimensionful constant $\chi$\footnote{In Ref.~\cite{Mirzagholi:2014ifa}, the parameter $\chi$ was promoted to a function of the field $\varphi$. This was proved crucial for generating the required amount of IDM in the early Universe 
(the issue analogous to getting the correct relic abundance of DM in the CDM framework). On the other hand, the slight dependence on the field $\varphi$ is completely irrelevant for the discussion of the present paper, where we focus 
on the Solar system scales.}, IDM reduces to a pressureless perfect fluid, with 
the fields $\Sigma$ and $\varphi$ playing the role of its energy density and velocity potential, respectively. Switching on the higher derivative term renders the fluid slightly imperfect~\cite{Mirzagholi:2014ifa} (hence, 
the name) and equips it with a non-zero sound speed $c_s \sim \frac{\sqrt{\chi}}{M_{Pl}}$~\cite{Chamseddine:2014vna}. In what follows, we assume no direct coupling between IDM 
fields and the standard matter fields. This guarantees that IDM may constitute the invisible matter in the Universe, at least for not large values of $c_s$.

In the synchronous gauge, a homogeneous solution for the field $\varphi$ takes a simple form,
\begin{equation}
\label{synchr}
{\bar \varphi}=t \; .
\end{equation}
So, it serves as the time parametrization.  Note an unconventional dimensionality $-1$ of the scalar $\varphi$.  Existence of the preferred frame, i.e., the one, where the background value of the field ${\varphi}$ is given by Eq.~\eqref{synchr}, 
implies dynamical Lorentz violation in the model~\eqref{action}. In this regard, it is akin to the Einstein--Aether theory~\cite{Jacobson:2000xp}\footnote{For this reason, the model~\eqref{action} is referred to as the 'scalar Einstein--Aether' in Ref.~\cite{Haghani:2014ita}.}. The latter, however, deals with a unit 4-vector field $u_{\mu}$ rather than with the 
4-gradient of a scalar. This distinction produces drastically different dynamics in the two models.

Particularities of the cosmological evolution in the scenario~\eqref{action} have been considered 
in Refs.~\cite{Capela:2014xta, Mirzagholi:2014ifa, Ramazanov:2015pha}. The main effect stems from non-zero sound speed~\cite{Chamseddine:2014vna}. Namely, the formation of objects with the 
size smaller than the sound speed horizon is suppressed compared to the predictions of Cold DM (CDM). Consequently, one risks 
to strongly affect the bottom-up picture of the structure formation for sufficiently large values of the parameter $\chi$. In Ref.~\cite{Capela:2014xta}, this observation was used to set the bound, 
\begin{equation}
\label{limitstructure}
\frac{\chi}{M^2_{Pl}} \lesssim 10^{-10} \; .
\end{equation}
For much smaller values of the parameter $\chi$, IDM is indistinguishable from CDM at the cosmological 
level. On the other hand, given values saturating the bound in Eq.~\eqref{limitstructure}, the behavior of IDM shares some similarities with the 
Warm DM. 

Let us briefly discuss the quantum features of the model~\eqref{action}. Compared 
to GR, it propagates three degrees of freedom: the standard ones associated with two 
polarizations of the helicity-2 graviton, and the scalar potential $\Psi$, which is now a dynamical 
field~\cite{Horava:2009uw, Sotiriou:2009gy, Sotiriou:2009bx,  Koyama:2009hc, Blas:2010hb, Ramazanov:2016xhp}\footnote{These references deal with the projectable Horava--Lifshitz 
model, which nevertheless reproduces the action~\eqref{action} in the certain limit. See the discussion below.}. The extra degree of freedom $\Psi$ exhibits gradient/ghost instabilities for negative/positive values of the parameter $\chi$. 
This, however, does not invalidate the model immediately. Indeed, those pathologies are not particularly dangerous 
provided that there is a sufficiently low Lorentz-violating cutoff on the spatial momenta of the modes of the field $\Psi$~\cite{Cline:2003gs, Rubakov:2008nh}. That cutoff 
is associated with the scale of yet unknown UV completion of the model or the strong coupling scale. 

In the version  with gradient 
instabilities ($\chi <0$), however, this cutoff turns out to be extremely low~\cite{Blas:2010hb}, $\Lambda \ll (0.1~\mbox{mm})^{-1}$. The latter contradicts the tests of gravity extending 
from the sub-mm distances to the Solar system scales. On the other hand, the ghost unstable 
branch of the model ($\chi >0$) allows for the cutoff scale $\Lambda$ as large as $\Lambda \sim 10$ TeV. Assuming that the 
strong coupling and UV scales are of the same order, this bound implies the 
constraint on the parameter $\chi$~\cite{Ramazanov:2016xhp}, 
\begin{equation}
\label{micros}
\frac{\chi}{M^2_{Pl}} \lesssim 10^{-20} \; .
\end{equation}
For larger values of the constant $\chi$, the vacuum decay with photons 
and ghosts in the final state is too fast, what leads to the conflict 
with measured fluxes of the gamma- and X-ray emission~\cite{Sreekumar:1997un}.

One interesting way of UV completing the action~\eqref{action} is suggested  in the context of the projectable Horava--Lifshitz model~\cite{Horava:2009uw}. The latter postulates 
a non-uniform transformation of time and spatial 
coordinates under the scaling. This has a dramatic effect on the ultraviolet behaviour of 
gravitons resulting into the strong distortion of their dispersion relation, $\omega^2 \sim {\bf p}^6$. 
Consequently, there are less divergences in the graviton loop integrals, what eventually leads to the (power counting) renormalizability of gravity~\cite{Horava:2009uw, Barvinsky:2015kil}. While the Horava--Lifshitz model is 
manifestly non-relativistic, it allows for the covariant description by introducing 
the St$\ddot{\mbox{u}}$ckelberg field $\varphi$ dubbed khronon. Then, its infrared limit exactly takes the form~\eqref{action}~\cite{Blas:2010hb, Blas:2009yd}. It is thus not a surprise that DM has been identified in this context~\cite{Mukohyama:2009mz}.
 In particular, the 
term with the Lagrange multiplier ensures the projectablility condition, which eliminates 
the pathological mode otherwise present at low momenta~\cite{Blas:2009yd, Charmousis:2009tc}\footnote{Alternative ways to tackle the pathological mode, not invoking for the projectability condition, are possible and 
have been proposed in Refs.~\cite{Blas:2009qj, Horava:2010zj}.}. The parameter $\chi$ is generically non-zero\footnote{The notation $\chi$ may be inconvenient 
for those familiar with Horava--Lifshitz gravity. There one deals with the parameter $\lambda$, which appears in front of the term $\sim K^2$ (the trace of the extrinsic curvature tensor squared) in the ADM formulation 
of the model. The two are related to each other by $\chi=\frac{1-\lambda}{8\pi G}$.}, and is supposed to 
follow the renormalization group flow towards the 'GR point' $\chi =0$.

Embedding the model~\eqref{action} into the Horava--Lifshitz gravity is not without 
problems, though. The reason is that UV operators modifying the dispersion 
relation of the gravitons are not capable of curing ghost instabilities. 
Hence, the only way to stabilize the catastrophic vacuum decay 
is to assume that the model enters a putative strong coupling phase. This severely 
obstructs the main objective of the Horava's proposal---perturbative 
renormalization of gravity.

Recently, the action~\eqref{action} has been rediscovered in a completely different framework of the 
mimetic matter~\cite{Chamseddine:2014vna, Chamseddine:2013kea}. There one deals with a non-invertible conformal transformation of the 
metric~\cite{Deruelle:2014zza}, 
\begin{equation}
\nonumber 
\tilde{g}_{\mu \nu}=A(\varphi, X) g_{\mu \nu} +B(\varphi, X) \partial_{\mu} \varphi \partial_{\nu} \varphi \; ,
\end{equation}
where $A$ and $B$ are the arbitrary functions of the scalar $\varphi$ and $X \equiv g_{\mu \nu} \partial^{\mu} \varphi \partial^{\nu} \varphi$. 
Performing this transformation on the standard Einstein--Hilbert action, one 
does not reproduce GR, but rather GR supplemented by a perfect pressureless fluid dubbed mimetic DM. Equivalently, the latter can be introduced by making use of the Lagrange 
multiplier as in Eq.~\eqref{action}, i.e., without directly referring to the 
disformally transformed metric~\cite{Golovnev:2013jxa, Hammer:2015pcx}. The higher derivative term is absent in the original formulation 
of the mimetic matter scenario. Nevertheless, that extension does not affect the main idea underlying the scenario, and even appears to be the only viable option for a number of 
phenomenological issues~\cite{Chamseddine:2014vna, Capela:2014xta}\footnote{Different extensions have been considered in Refs.~\cite{Chamseddine:2014vna, Arroja:2015wpa, Arroja:2015yvd}. 
The former equips the field $\varphi$ with some potential $V(\varphi)$. In this way, one manages to mimic fairly arbitrary cosmological evolution. In Refs.~\cite{Arroja:2015wpa, Arroja:2015yvd}, the mimetic matter scenario 
has been extended by means of the Horndeski higher derivative terms.}.

In the present paper, we do not assume any particular gravitational framework behind 
the model~\eqref{action}. Neither, we are interested in its cosmological or microscopic manifestations. Our main focus here is the  intermediate range of scales: Solar system. Surprisingly, this yields 
limits, which are many orders of magnitude stronger than those obtained 
from the structure formation considerations. Moreover, our discussion does not assume that IDM gives the dominant contribution to the invisible matter in the Universe (namely, IDM may constitute only its tiny fraction).

The behavior of IDM in the Solar system shares some features (but not all) with CDM. Let us briefly summarize the main effect. 
The Sun moves relative to the cosmic microwave background and Milky Way rest frames with the speed $v \simeq 10^{-3}$.  Naturally, the preferred frame is associated with one of those. 
An observer in the Solar system sees a flow of IDM. Affected by the gravitational potential of the Sun, 
the flow focuses downstream from the Sun forming a caustic. This part of the story parallels to that of CDM. 

In the latter case, however, gravitational focusing is a rather moderate effect 
introducing a few percent correction to the annual 
modulation of the flux of  DM particles passing through the Earth~\cite{Danby, Danby1, Griest:1987vc,  Sikivie:2002bj, Belotsky1, Belotsky2, Lee:2013wza, Patla:2013vza}. 
The things are different in the IDM scenario. The reason is the higher derivative structure of its action~\eqref{action}. 
The term $\sim \chi (\square \varphi )^2$ serves as a powerful source of the IDM energy density $\Sigma$. Recall that IDM is not coupled directly to the standard matter, but only gravitationally. 
Therefore, the amplification of the field $\Sigma$ has no consequences for particle experiment facilities. On the flipside, the field $\Sigma$ backreacts on the space-time geometry causing distortion of the metric. This 
is the main effect identified in the present paper\footnote{On the contrary, in the case of CDM the distortion of the metric is a negligible effect, while the predictions for the particle exxperiments can be sensisble.}. As in the case of CDM, it is particularly prominent in the direction opposite to the velocity of the Sun ${\bf v}$, where IDM is mainly accumulated. Borrowing terminology of Ref.~\cite{Dubovsky:2004qe}, astrophysical objects moving relative to the 
preferred frame~\eqref{synchr} leave  'star tracks' behind them. 

Remarkably, in the IDM scenario, deviations from the GR metric emerge only in the second order in the Newtonian potential. 
This is in a sharp contrast to the predictions of other models, which deal with preferred frame effects~\cite{Blas:2010hb, Foster:2005dk, Blas:2014aca}. In our case, the latter manifest via the spatial dependence of coefficients measuring 
quadratic corrections to the GR metric. Such a correction does not exactly follow the standard parametrized post-Newtonian (PPN) approach, where analogous coefficients are assumed to be constant. 
In particular, the parameter $\beta$ akin to the PPN one $\beta_{PPN}$, deviates from unity by $\beta -1\simeq \frac{4\pi \chi}{M^2_{Pl} v^4} \cdot \frac{1}{\theta^4}$, where 
$\theta$ is the angle between the line of sight of the observer on the Sun and the direction $-{\bf v}$. Ignoring $\theta$-dependence here, we convert bounds on $\beta_{PPN}$ following from the studies 
of the Mercury perihelion precession into the limit 
on the theory constant $\chi$. The resulting constraint is quite stringent: $\chi/M^2_{Pl} \lesssim  10^{-18}$, which is only two orders of magnitude weaker than the limit~\eqref{micros} inferred from the microscopic physics considerations. 
Notably, this is a conservative constraint, since neglecting the $\theta$-dependence we underestimate the actual effect due to gravitational focusing.

The remainder of the paper is as follows. In Section~2, we deduce equations of motion following from the action~\eqref{action}. We discuss the main assumptions and approximations used to study the dynamics of IDM in Section~3. For the sake of 
convenience, there we also outline the main formulae describing the IDM profile in the Solar system, as well as the 
induced metric corrections. Derivation of those results is explained in Sections~4 and~5, where we restrict to the linear and quadratic order analysis in the Newtonian potential, respectively. 
In Section~4, we also identify the narrow region of space, where the perturbative description of IDM breaks down. We assess metric perturbations 
in this region in Section~6. The reader interested  in the final results, may skip Sections 4-6 and go directly to Section 7, where we contrast our predictions for metric perturbations to the 
observational bounds, and derive the constraint on the theory constant $\chi$.

\section{Setup}

Let us write down equations of motion following from the action~\eqref{action}. Variation with respect to the Lagrange multiplier $\Sigma$ yields
\begin{equation}
\label{constraint}
g_{\mu \nu} \partial^{\mu} \varphi \partial^{\nu} \varphi=1 \; .
\end{equation}
Applying the covariant derivative to both parts of the constraint, one 
reproduces the geodesics equation followed by test particles in the gravitational field~\cite{Blas:2009yd, Lim:2010yk}, 
\begin{equation}
\label{geodesics}
\frac{du^{\mu}}{ds} +\Gamma^{\mu}_{\nu \lambda} u^{\nu} u^{\lambda}=0 \; ,
\end{equation}
where $u^{\mu} \equiv \partial^{\mu} \varphi$ is the 4-velocity. In the following, we will repeatedly make use of this 
analogy with the case of dust particles.

Varying the action~\eqref{action} with respect to the field $\varphi$, we obtain
\begin{equation}
\label{conservation}
\nabla_{\mu} \left( \Sigma \nabla^{\mu} \varphi \right)=\chi \square^2 \varphi \; .
\end{equation}
For the choice $\chi=0$, Eq.~(\ref{conservation}) takes the form of the energy density conservation for a pressureless perfect fluid. In this limit, there is no source of DM: if its density is zero everywhere 
at some moment, it remains so at later times. The degeneracy with the case of CDM gets broken upon switching on a non-zero
$\chi$. Then, any non-trivial configuration of the field $\varphi$ serves as the source of DM. 
This difference between IDM and CDM will be important for our further discussions. 

Finally, there are Einstein equations, which we write in the form, 
\begin{equation}
\label{Einstein}
R_{\mu \nu}= 8\pi G \left( T_{\mu \nu}-\frac{1}{2} Tg_{\mu \nu}\right) \; ,
\end{equation}
where 
\begin{equation}
\nonumber 
T_{\mu \nu}=T^{matter}_{\mu \nu}+T^{IDM}_{\mu \nu} \; .
\end{equation}
Here $T^{matter}_{\mu \nu}$ and $T^{IDM}_{\mu \nu}$ are the stress-energy tensors corresponding to the standard matter and IDM. The latter is  given by~\cite{Chamseddine:2014vna}
\begin{equation}
\label{mimstress}
T^{IDM}_{\mu \nu}=\Sigma \nabla_{\mu} \varphi \nabla_{\nu} \varphi +\chi \left(\nabla_{\alpha} \varphi \nabla^{\alpha} \square \varphi+\frac{1}{2} (\square \varphi)^2 \right) g_{\mu \nu}-\chi (\nabla_{\nu} \varphi
\nabla_{\mu} \square \varphi +\nabla_{\nu} \square \varphi \nabla_{\mu} \varphi) \; .
\end{equation}
In the limit $\chi \rightarrow 0$, this reduces to the stress-energy tensor of a pressureless perfect  fluid. 
In what follows, we will refer to the field $\Sigma$ as the energy density of IDM even for non-zero values of the parameter $\chi$. 
The discrepancy between the physical and so defined energy density is small and irrelevant for our study. 
This can be verified rigorously at each step of calculations.

\section{Flow of IDM past the Sun: assumptions and main formulae}

Our goal in the present paper is to study the imprint of IDM on the metric created by the 
standard matter---the Sun. For this purpose, we employ the test field approximation. 
Namely, we assume that the Sun is the main source of the space-time 
curvature, while IDM produces small corrections on the background metric. This assumption 
will be checked by the structure of the final result. 

It is convenient to perform calculations in the rest frame of the source. In that coordinate 
system, there is a flow of IDM moving towards the Sun with the velocity ${\bf v} \equiv v^i$. Performing the Lorentz transformation on the background profile~\eqref{synchr} of the field 
$\varphi$, we get
\begin{equation}
\label{backgroundvarphi}
\bar{\varphi}= \sqrt{1+v^2}t -v^i x^i \; .
\end{equation}
One may naively expect the speed $v \equiv |{\bf v}|$ to play a role of a small expansion parameter, when evaluating metric corrections. This is indeed the case of the Einstein--Aether or khronometric theories~\cite{Blas:2010hb, Foster:2005dk}. 
The situation is different in the IDM scenario: at least sufficiently far from the Sun, 
the velocity 
enters metric corrections as $1/v^n$, where $n$ is some positive number. This peculiar dependence 
on the quantity $v$ has a clear physical explanation. If IDM passes fast near the Sun, the latter does not have enough time 
to affect the flow appreciably. Hence, the effects of gravitational focusing milden in that case.

\begin{figure}[tb!]
\begin{center}
\includegraphics[width=0.90\columnwidth,angle=0]{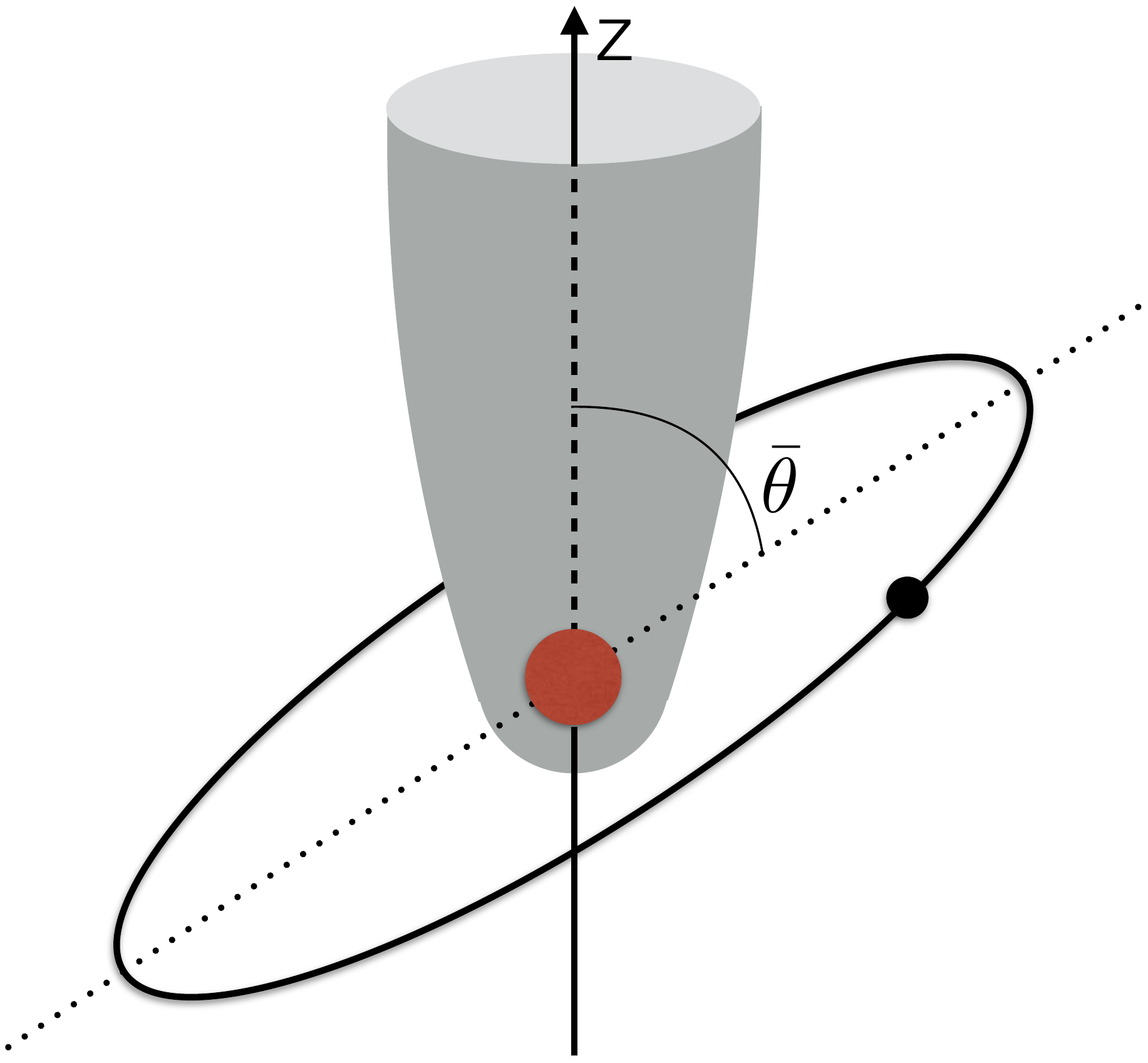}
\end{center}
\caption{The ecliptic plane is shown relative to the direction of the IDM flow aligned with the $z$-axis. The shaded region corresponds to the part of the space, where the perturbative calculation of the 
IDM profile and the induced metric corrections breaks down. This covers not only the region downstream from the Sun, but also some small region near its surface.}\label{plot}
\end{figure}

The perturbative treatment of the metric is still possible in terms of the Newtonian potential, provided that the following 
inequality is obeyed, 
\begin{equation}
\label{inequality}
|\Phi| \ll v^2 \; .
\end{equation}
 The potential $\Phi$ is given by 
\begin{equation}
\nonumber 
\Phi=-\frac{M}{M^2_{Pl}r} 
\end{equation}
outside the source, where $M$ is the Sun's mass.
In other words, the actual expansion parameter is $\frac{|\Phi|}{v^2}$, and we anticipate linear and quadratic  metric corrections to be of the order $\frac{\chi}{M^2_{Pl}} \cdot \frac{\Phi}{v^2}$ and $\frac{\chi}{M^2_{Pl}} \cdot \frac{\Phi^2}{v^4}$, 
respectively. This fact places limitations on the analysis of the present paper: it is not applied for sufficiently strong gravitational fields/small speeds. On the other hand, for the choice 
\begin{equation}
\label{velocity}
v \simeq 10^{-3} \; ,
\end{equation}
the inequality~\eqref{inequality} 
represents a reasonable approximation within the Solar system. So, one has $\frac{|\Phi|}{v^2} = \frac{v^2_{Mercury}}{v^2} \simeq \frac{1}{36}$ at the Mercury distance, 
where $v_{Mercury} \approx 47~\mbox{km/s}$ denotes its average orbital speed. For more remote planets such a ratio decreases gradually, increasing the accuracy of the inequality~\eqref{cond}. 
The only region of space, where Eq.~\eqref{inequality} is not valid, is very close to the surface of the Sun, see Fig.~\ref{plot}. Being primarily interested in the planetary dynamics, we will ignore this 
subtlety in the bulk of the paper. The issue, however, comes into play, when calculating metric corrections induced by IDM. To fill in the gap, 
in Appendix~A we provide a simplified analysis of the relevant fields in this region. 

Before going into the details of the analysis, let us briefly comment on the origin of the estimate~\eqref{velocity}. It is natural to associate the preferred frame with the rest frame either of the cosmic microwave background or the Milky Way halo.  
In the former case, the speed is inferred from the observed dipole anisotropy~\cite{planck} and is given by $v \approx 369~\mbox{km/s}$. 
In the latter case, $v \approx 220~\mbox{km/s}$. We see that both match the estimate~\eqref{velocity} 
very well. Still 
the two choices are not degenerate, as they imply different inclinations of the velocity ${\bf v}$ relative to the Solar 
system plane. This distinction will be relevant, when contrasting our theoretical predictions to the 
observational data.

When defining metric corrections, our strategy will be as follows. First, we will solve the constraint equation~\eqref{constraint} perturbatively expanding the 
field $\varphi$ in the powers of the 
Newtonian potential. Hereafter, we assume a {\it stationary} flow of IDM and, henceforth, restrict to a static configuration of the velocity potential $\varphi$. In this approximation, the field $\varphi$ is uniquely defined by the boundary condition
\begin{equation}
\label{boundary}
{\nabla  \varphi} \rightarrow -{\bf v}~~~\mbox{at}~~~z \rightarrow -\infty \; .
\end{equation}
We choose the $z$-axis of the reference frame to be aligned with the velocity ${\bf v}=v^i$, and set its origin at the location of the Sun, as shown in Fig.~\ref{plot}. 
Downstream 
from the Sun $(z>0)$, the perturbative series of interest is schematically given by
\begin{equation}
\label{schematically}
\delta \varphi =\eta_1 \frac{M}{vM^2_{Pl}} \ln \frac{2z \rho_0}{\rho^2} +\eta_2 \frac{M^2}{v^3 M^4_{Pl}} \frac{z}{\rho^2}+
{\cal O} \left( \frac{M^3}{v^5M^6_{Pl}} \cdot \frac{z^2}{\rho^4}\right) \; ,
\end{equation}
(see Eqs.~\eqref{sollinear} and~\eqref{varphi2gen} for the exact expressions) where $\delta \varphi \equiv \varphi-\bar{\varphi}$. Here $\eta_1$ and $\eta_2$ are 
order one coefficients, and $\rho_0$ is an irrelevant dimensionful constant. 
Now, we see explicitly the relevance of the condition~\eqref{inequality}---had it not been obeyed, the 
series~\eqref{schematically} would not converge. Still, the terms entering the series~\eqref{schematically} formally 
blow up in the limit $\rho /z \rightarrow 0$, what eventually invalidates the perturbative treatment of the velocity potential $\varphi$, see Fig.~\ref{plot}. Fortunately, there is a way to find a solution for the field $\varphi$  in this narrow region. 
As we will see in Section~6, the role of non-perturbative effects is to smoothen the singularities present in Eq.~\eqref{schematically}, so that the actual field $\varphi$ is finite everywhere. 
On the other hand, there is no such an issue for negative $z$ (the upstream region). 
In that case, the perturbative series analogous to Eq.~\eqref{schematically} remains well-defined for arbitrarily small values of $\rho$.

As for the next step, we calculate the energy density $\Sigma$. The latter is mainly fixed by the configuration of the field $\varphi$, while the metric perturbations by themselves give a negligible contribution at this level. 
Just like the velocity potential $\varphi$, the density field $\Sigma$ can be treated perturbatively almost in the whole region of space. Cutting the series at the quadratic term, one has for $z>0$
\begin{equation}
\label{sigmaschematically}
\Sigma \simeq \frac{\chi M^2}{v^4 M^4_{Pl}} \cdot \frac{z^2}{\rho^6} +{\cal O} \left(\frac{\chi M^3}{v^6 M^6_{Pl}} \cdot \frac{z^3}{\rho^8} \right) \; ,
\end{equation}
(see Eqs.~\eqref{sigmalinsun} and~\eqref{sigma2} for the exact expressions). We neglected here the background value of the field $\Sigma$. This can be easily justified. Indeed, the amount of IDM generated by the gravitational field of 
the Sun turns out to be much larger than its 
expectation value in the surrounding halo. Furthermore, IDM is not produced at the linear level. We will show this explicitly in Section~4. Once again, the sharp $\rho$-dependence of the energy density in the limit $\rho \rightarrow 0$ is an artefact of the 
perturbative approach, which breaks down near the $z$-axis. At the same time, the non-perturbative effects  convert the strong growth of the 
field $\Sigma \sim \frac{1}{\rho^6}$ as in Eq.~\eqref{sigmaschematically} into a milder one $\Sigma \sim \frac{1}{\rho^2}$. See Section~6 for details.

 Knowing the distribution of the fields $\Sigma$ and $\varphi$, we solve the Einstein equations and obtain metric perturbations defined as
\begin{equation}
\label{definition}
h_{00} \equiv g_{00}-1 \qquad h_{0i} \equiv g_{0i} \qquad h_{ij} \equiv g_{ij}+\delta_{ij} \; . 
\end{equation}
There is a subtlety at this point. While the fields $\Sigma$ and $\varphi$ can be calculated unambiguously in the perturbative region, to find 
the metric perturbations one has to know the IDM profile everywhere in space, including non-perturbative region. To tackle this issue, we extrapolate 
results of the perturbative analysis for the fields $\Sigma$ and $\varphi$ from the region, where this calculation can be trusted, to the whole space. We also approximate the 
Sun by a pressureless perfect fluid and set to zero relative motion of its constituents. It is then 
straightforward to evaluate the metric. To the quadratic order, the correction to its $00$-component reads
\begin{equation}
\label{00schematically}
\delta h_{00} \simeq \frac{\chi}{M^2_{Pl}} \cdot \frac{\Phi^2}{v^4} \cdot \frac{1}{\theta^4} \; ,
\end{equation}
(see Eq.~\eqref{002} for the exact expression) where $\theta$ is the angle between the line of sight of the observer sitting on the Sun and the direction ${\bf v}$. The absence of linear order corrections here 
reflects the fact that the perturbative expansion~\eqref{sigmaschematically} for the Lagrange multiplier field $\Sigma$ starts from the quadratic term. The same conclusion holds for the $0i$- and $ij$-metric components. 
Strictly speaking, metric perturbations calculated in this way, at best may serve as the estimate for the true ones. Therefore, we should critically assess the discrepancy 
caused by the non-perturbative effects. This is done in Section~6, where we discuss the IDM profile downstream from the Sun. In  Appendix~A, we estimate 
non-perturbative effects due to the breakdown of the inequality~\eqref{cond}, which occurs near its surface. In both cases the discrepancy 
is not degenerate with the result~\eqref{00schematically} by the magnitude or by the spatial dependence. 
Hence, we can separate the effects encoded in Eq.~\eqref{00schematically} from those following from the non-perturbative 
distribution of the IDM profile.

With this said, we contrast the resulting metric correction~\eqref{00schematically} to the existing experimental bounds on the 
deviations from GR manifested as the constraints on the PPN parameters.  There is subtlety, however. 
One of the prescriptions underlying the PPN approach is the smooth distribution of the fields, which source the 
space-time curvature~\cite{Will:2014kxa}. This is why it cannot be applied directly to the model under study. First, the fields $\varphi$ and $\Sigma$ vary fast in the narrow region downstream from the Sun, what eventually leads to the 
breakdown of the perturbative approach. 
Even away from this part of space, the $\theta$-dependence of the metric correction~\eqref{00schematically} does not allow for the direct comparison with its PPN counterpart governed by the {\it constant} parameter 
$\beta_{PPN}$. It is, however, still possible to impose a conservative limit on the theory constant $\chi$. 
Indeed, the metric correction~\eqref{00schematically} has a non-vanishing minimal value with respect to the angle $\theta$. Note that for a fixed angle $\theta$, Eq.~\eqref{00schematically} exactly takes the form 
as in the PPN formalism. Then, the model can be constrained using the existing bounds on the parameter $\beta_{PPN}$. See Section~7 for more details.

Our last, but not the least important, assumption concerns the behavior of IDM in the deeply non-linear regime, where a caustic is supposed to be formed. 
This issue is also closely related with one of our basic approximations---the stationarity of IDM flow. While the aforementioned non-perturbative effects 
substantially smoothen the IDM profile downstream from the Sun, singularities are not completely avoided\footnote{In particular, the energy density $\Sigma$ and the metric correction $\delta h_{00}$ 
grow as $\Sigma \sim \frac{1}{\rho^2}$ and $\delta h_{00} \sim \ln^2 \theta$ in the non-perturbative regime, respectively. See the discussion after Eq.~\eqref{sigmaschematically} and in Section~6.}. 
The reason for this is the presence of the 
constraint~\eqref{constraint}, which effectively describes the potential flow of dust particles. Note that the solution for the field $\varphi$ is defined by Eq.~\eqref{constraint}, while the higher derivative term in the action~\eqref{action} is irrelevant at this level.
Therefore, the analogy with the CDM case indeed works. It is well-known that  particles following the geodesics equation~\eqref{geodesics} eventually cross at what is called caustic, where the physical 
quantities, i.e., the energy density and the divergence of the velocity, blow up. In the CDM framework, however, 
the appearance of caustics merely hints the breakdown of a single flow approach for a collection of particles. Since then, one should switch to the multi-flow description. On the contrary, IDM is fundamentally 
described by a single flow, parametrized in terms of the unique field $\varphi$. Therefore, an appearance of a caustic may imply a fundamental problem of the theory.
On the other hand, formation of caustics is generic in all scalar-tensor theories, even in models which do not exhibit a dust-like behavior~\cite{Babichev:2016hys}.

Mechanisms for avoiding the caustic singularity in the model at hand 
have been discussed in Refs.~\cite{Capela:2014xta, Mukohyama:2009tp}. So, Ref.~\cite{Capela:2014xta} suggests that the test field approximation is exceeded at some point, and the field $\Sigma$  becomes the main source 
of gravity. It is then crucial that the energy density $\Sigma$ may take negative values, so that the gravitational force becomes repulsive. In this situation, 
there is the chance to avoid the caustic singularity. 
Alternatively, anti-gravity is produced for a particular choice of UV operators completing the model~\eqref{action}, 
as it has been argued in Ref.~\cite{Mukohyama:2009tp} in the Horava--Lifshitz framework. We assume in what follows that these mechanisms of smoothening the singularity are compatible with the stationarity of the IDM flow\footnote{See Refs.~\cite{Izumi:2009ry, Greenwald:2009kp} for the debates on the similar issues.}, at least at the scales and time intervals relevant for the discussion of the 
present paper.

\section{Linear level analysis}

In the bulk of the paper, we will evaluate the field $\delta \varphi \equiv \varphi -\bar{\varphi}$  in the straightforward 
manner---by resolving the constraint~\eqref{constraint}. Alternatively, one observes that Eq.~\eqref{constraint} leads to the geodesics 
equation~\eqref{geodesics} followed by test particles in the gravitational field. Thus, not resorting to Eq.~\eqref{constraint}, we could equivalently consider 
a parallel flow of dust particles falling onto the Sun with the velocity ${\bf v}$. The velocity distribution, which 
characterizes the flow, should match the one obtained directly from Eq.~\eqref{constraint}. 
This will serve as the cross-check of our computations. We relegate the further details to Appendix~B.

We use the Newtonian gauge to fix the metric potentials induced by the standard matter. Then, outside of the Sun, the metric takes the form 
\begin{equation}
\label{metrictest}
ds^2=(1+2\Phi) dt^2 -(1-2\Phi) d{\bf x}^2 \; .
\end{equation}
In the stationary flow approximation, the constraint~\eqref{constraint} linearized reads  
\begin{equation}
\label{lin}
\partial_i \delta \varphi^{(1)} v^i =(1+2v^2)\Phi \; . 
\end{equation} 
Recall that we work in the rest frame of the Sun, where the background value of the field $\varphi$ is given by Eq.~\eqref{backgroundvarphi}. The general solution of Eq.~\eqref{lin} is given by
\begin{equation}
\label{sollin}
\delta \varphi^{(1)} =    -(1+2v^2)\frac{M}{v \cdot M^2_{Pl}} \mbox{arcsinh} \frac{z}{\rho} +f(\rho)\; ,
\end{equation}
where $f(\rho)$ is some function of integration. We rotated the reference frame in a way that its 
$z$-axis is aligned with the velocity ${\bf v}$, and the origin is placed at the location of the Sun, as shown in Fig.~\ref{plot}. The function $f(\rho)$ is uniquely defined from the boundary condition chosen by the analogy with the particle case, 
\begin{equation}
\label{boundarydelta}
{\bf \nabla} \delta \varphi \rightarrow 0~~~ \mbox{at}~~~ z \rightarrow -\infty \; ,
\end{equation}
which reflects Eq.~\eqref{boundary}. The first order solution for the field $\delta \varphi$ reads  
\begin{equation}
\label{sollinear}
\delta \varphi^{(1)}=- (1+2v^2)\frac{M}{vM^2_{Pl}} \ln \left[\frac{\rho_0}{\sqrt{z^2+\rho^2}-z} \right]\; ,
\end{equation}
where $\rho_0$ is an irrelevant dimensionful constant. Here we made use of an identity $\mbox{arcsinh} \frac{z}{\rho}=\ln \left[\frac{z}{\rho} +\sqrt{\frac{z^2}{\rho^2}+1}\right]$. This solution exhibits a regular behavior in the upstream region, 
\begin{equation}
\nonumber 
\delta \varphi^{(1)} =- (1+2v^2)  \cdot \frac{M}{vM^2_{Pl}} \ln \frac{\rho_0}{2|z|} \qquad \left( \frac{\rho}{|z|} \ll 1,~z<0 \right)   \; ,
\end{equation}
as it should be. A different story occurs for $z>0$, where the solution grows infinitely in the limit $\rho \rightarrow 0$ eventually exceeding the regime of the applicability 
of the linear approximation (see the discussion below), 
\begin{equation}
\nonumber 
\delta \varphi^{(1)}= -(1+2v^2) \cdot \frac{M}{vM^2_{Pl}} \cdot \ln \frac{2z \rho_0}{\rho^2}  \qquad \left( \frac{\rho}{z} \ll 1,~z>0 \right) \; .
\end{equation}
That configuration of the field $\delta \varphi$ has a clear physical meaning: it reflects the gravitational focusing experienced by IDM passing in the vicinity of the 
Sun. In the first order, however, this amplification of the field $\delta \varphi$ produces essentially no effect on the metric perturbations, as we will see below. 

When deriving Eq.~\eqref{sollinear}, we implied a point source approximation for the standard matter. It is 
straightforward to generalize it to the realistic case of a finite size source. The 'true' solution coincides with the one of Eq.~\eqref{sollinear} apart from the cylindric  region of the size $\rho=R$ along the direction ${\bf v}$, where 
$R$ is the radius of the Sun. This is, however, a minor issue. That is, the point source approximation is applied wherever the perturbative description of the field $\delta \varphi$ can be trusted\footnote{This is true for the gravitational field of the Sun, or more generally, 
a sufficiently compact object. On the other hand, for the light source like the Earth, the linear approximation is essentially never exceeded. We discuss this case in Appendix~C.}.

The inequality~\eqref{inequality}---one of the prerequisite for the perturbative expansion---breaks down near the surface of the Sun. 
 Furthermore, the linear approximation holds, only if the following conditions are obeyed,  
\begin{equation}
\nonumber
|\partial_z \delta \varphi | \ll v,~~~ \left( \partial_{\rho} \delta \varphi \right)^2 \ll |v\partial_z \delta \varphi| \; .
\end{equation}
Otherwise,  Eq.~\eqref{lin} gets modified due to the appearance of quadratic terms $\sim \partial_i \delta \varphi \partial_i \delta \varphi$. These translate into the inequality
\begin{equation}
\label{cond}
 \frac{\rho^2}{z^2}   \gg \frac{|\Phi |}{v^2} \qquad z>0 \; .
\end{equation}
Several remarks are in order here. First, it is worth 
pointing out that the restriction~\eqref{cond} concerns only the downstream region ($z>0$), while for $z<0$ the linear approximation is retained for arbitrary $\rho$. 
Second, for very small ratios $\rho/z$ not only the linear analysis, but the perturbative description of the field $\delta \varphi$ is questionable. 
We further clarify this point in Section~6. In particular, at $\rho/z \sim \sqrt{|\Phi|/v^2}$, the higher order corrections to the field $\delta \varphi$ all become of the same order, 
$ |\delta \varphi^{(1)} | \sim |\delta \varphi^{(2)}|  \sim ... \sim   |\delta \varphi^{(n)}|$. Instead, for larger ratios $\rho/z$  they are arranged in a well-defined perturbative series 
over the powers of the potential $\Phi$. This comment is not trivial, since only the first order quantity $\delta \varphi^{(1)}$ was used to derive the condition~\eqref{cond}. Therefore, 
one should check rigorously, if Eq.~\eqref{cond} is robust against including the higher order corrections into the analysis. For a while, we take this statement for granted, 
and give an explicit proof in Section~6.

Now, let us switch to the calculation of the Lagrange multiplier field $\Sigma$. We assume that it vanishes at the background level, i.e., $\bar{\Sigma}=0$, what is compatible with the configuration~\eqref{backgroundvarphi} 
of the field $\varphi$. In the linear order, the relevant part of the 
source term for the field $\Sigma$ (the r.h.s. of Eq.~\eqref{conservation}) can be written as follows, 
\begin{equation}
\nonumber 
(\square^ 2 \varphi )^{(1)}=-\Delta (\square \varphi )^{(1)} \; . 
\end{equation}
The generic expression for the quantity $\square \varphi$ in the static field approximation is given by
\begin{equation}
\label{square}
\square \varphi =\frac{1}{\sqrt{-g}} \partial_i \left(-\sqrt{-g} g^{ij}v^j +\sqrt{-g} g^{ij}\partial_j \delta \varphi +\sqrt{-g} g^{0i} \sqrt{1+v^2}\right) \; .
\end{equation}
Keeping only linear terms here, we obtain
\begin{equation}
\nonumber 
\left( \square \varphi \right)^{(1)} =-\Delta \delta  \varphi^{(1)} \; ,
\end{equation}
where $\Delta$ is the Laplacian understood with respect to the Euclidian metric. The shortcut way of calculating the quantity $\Delta \delta \varphi^{(1)}$ is to apply the Laplace operator to both parts of Eq.~\eqref{lin} and make use of the 
boundary condition~\eqref{boundarydelta}. The result is trivially zero wherever the perturbative 
analysis can be trusted, i.e., the inequalities~\eqref{inequality} and~\eqref{cond} are satisfied, 
\begin{equation}
\label{lindeltasun}
\Delta \delta \varphi^{(1)}= 0 \; .
\end{equation} 
Hence, there is no source of IDM energy density $\Sigma$ to the linear order, namely,
\begin{equation}
\label{sigmalinsun}
\Sigma^{(1)}=0 \; .
\end{equation}
Indeed, in the stationary flow approximation, the field $\Sigma$ is defined from the following equation, 
\begin{equation}
\label{sigma1}
\partial_i \left(\Sigma^{(1)} v^i \right)=\chi \Delta^2 \delta \varphi^{(1)}\; .
\end{equation}
We assume vanishing of the field $\Sigma$ at $z \rightarrow -\infty$ as for the boundary 
condition. In other words, we neglect the fact that the Solar system is surrounded by 
the DM halo fully or partially made of IDM. This uniquely fixes the solution~\eqref{sigmalinsun} for the energy density $\Sigma^{(1)}$. As it follows, the stress-energy tensor of IDM~\eqref{mimstress} vanishes to the linear order. 
Hence, the metric~\eqref{metrictest} does not receive any corrections at this level.

This result, as we mentioned before, was obtained upon extrapolating the expressions~\eqref{lindeltasun} and~\eqref{sigmalinsun} to the whole space. Generically, 
metric perturbations calculated in that way are defined modulo the solution $h^{bound}_{\mu \nu}$, which 
satisfies the homogeneous Einstein equations $R_{\mu \nu}=0$ in the perturbative region~\eqref{cond}. The contribution $h^{bound}_{\mu \nu}$ is sourced by the 
IDM distribution in the narrow boundary region $\frac{\rho^2}{z^2} \lesssim \frac{|\Phi|}{v^2}$ (hence the subscript 'bound'). We estimate it in Section~6, as soon as the 
full IDM profile becomes available.

Our last comment is in order here. One could be interested in the 
situation, when the linear approximation holds essentially in the 
whole space. This is a realistic situation for a sufficiently light 
object, like the Earth. We relegate details of calculations to Appendix~C. Here let us quote the main result: modulo a gauge transformation, metric corrections are still zero in that case. 
Though this conclusion is not directly applied to the case of the Sun, the indication is clear: one should not expect any 
substantial deviations from GR to the linear order.

\section{Second order analysis}

In the previous Section, we observed that gravitational focusing of IDM 
is irrelevant to the first order. Namely, it does not lead to the production of the 
field $\Sigma$. A different story occurs in the second perturbation order,---the main subject of the present Section. To simplify calculations, hereafter we keep only the terms with the largest power of the quantity $1/v$.

In this approximation, the equation defining the quadratic correction to the field $\varphi$ is given by 
\begin{equation}
\label{varphi2equa}
 -2 \partial_i \delta \varphi^{(2)} v^i +\partial_i \delta \varphi^{(1)} \partial_i \delta \varphi^{(1)}=0 \; ,
\end{equation}
where we again assumed the stationarity of IDM flow. Here we omitted the terms arising from the quadratic order metric expansion, $\sim \Phi^2$, as well as the quantities $\sim v \Phi \partial \delta \varphi^{(1)}$. These are suppressed by the factor $\sim v^2$ compared 
to those present in Eq.~\eqref{varphi2equa}. 
We also made use of the test field approximation and neglected metric corrections induced by IDM. As it follows from Eq.~\eqref{varphi2equa}, 
metric perturbations do not directly affect the velocity potential in the second order.
The former enter only 
via the quantity $\delta \varphi^{(1)}$. This is a generic fact, which takes place for all the higher order corrections 
to the field $\varphi$.

Integrating out Eq.~\eqref{varphi2equa}, one obtains
\begin{equation}
\label{varphi2gen}
\delta \varphi^{(2)}= \frac{M^2}{v^3 M^4_{Pl}} \cdot \frac{1}{\sqrt{z^2+\rho^2}-z} \; .
\end{equation}
It is easy to see that the second order velocity perturbation, $-\nabla \delta \varphi^{(2)}$, goes to zero in the limit $z \rightarrow -\infty$, as it should be. 
In particular, for $z <0$, the field $\delta \varphi^{(2)}$ is independent of the variable $\rho$  in the limit $\rho \rightarrow 0$, 
\begin{equation}
\label{zless0}
\delta \varphi^{(2)}= -\frac{M^2 }{2v^3 M^4_{Pl} z}  \qquad \left( \frac{\rho}{|z|} \ll 1,~z<0 \right)\; ,
\end{equation} 
and thus exhibits the regular behavior upstream from the Sun. To the contrast, in the downstream region, the asymptotic behavior of the field $\delta \varphi^{(2)}$ is described by
\begin{equation}
\label{smallrhovelocity2}
\delta \varphi^{(2)} = \frac{2M^2}{v^3 M^4_{Pl}} \cdot \frac{z}{\rho^2} \qquad \left( \frac{\rho}{z} \ll 1, ~z >0 \right) \; .
\end{equation}
Namely, we gained the large factor $\sim \frac{z^2}{\rho^2}$ compared to Eq.~\eqref{zless0}. This amplification is exactly the manifestation of 
gravitational focusing. Moving towards very small values of the variable $\rho$, however, one subsequently breaks down the validity of the inequality~\eqref{cond}. In the non-perturbative phase, the 
result~\eqref{smallrhovelocity2} is not applied, and one should find another way to treat the field $\varphi$. We postpone this task until Section~6.  On the other hand, wherever the inequality~\eqref{cond} is obeyed, 
the second order quantity~\eqref{varphi2gen} is consistently smaller than the linear one~\eqref{sollinear}, $|\delta \varphi^{(2)}| \ll  |\delta \varphi^{(1)}|$. This justifies the statement made in Section~4. 

Now, let us evaluate the energy density $\Sigma$ in the second order. Again keeping only the leading powers of the quantity $1/v$, we write the relevant equation as
\begin{equation}
\label{eqdefsigma2}
\partial_i \left( \Sigma^{(2)} v^i \right)  =\chi \Delta^2 \delta \varphi^{(2)} \; .
\end{equation}
To the linear order, we remind, the energy density $\Sigma$ equals to zero. 
This explains the absence of the term $\sim \partial_i \left(\Sigma^{(1)} \partial_i \delta \varphi^{(1)} \right)$. Integrating out Eq.~\eqref{eqdefsigma2} with the boundary condition $\Sigma^{(2)} \rightarrow 0$ at $z \rightarrow -\infty$, one gets
\begin{equation}
\label{sigma2}
\Sigma^{(2)} =\frac{8\chi M^2}{v^4 M^4_{Pl}} \cdot \frac{1}{\sqrt{z^2+\rho^2}(\sqrt{z^2+\rho^2}-z)^3} \; .
\end{equation}
The latter exhibits a regular behavior for all $z<0$, while for $z>0$ there is a singularity in the formal limit $\rho \rightarrow 0$,
\begin{equation}
\label{sigmapertzplus}
\Sigma^{(2)} =\frac{64 \chi M^2}{v^4 M^4_{Pl}} \frac{z^2}{\rho^6} \qquad \left( \frac{\rho}{z} \ll 1, ~z >0 \right) \; .
\end{equation} 
This enhancement translates into the strong amplification of metric perturbations. Once again, such a strong dependence on the variable $\rho$ is trustworthy to the extent that the 
inequality~\eqref{cond} holds. Otherwise, non-perturbative effects cannot be ignored. These will be considered in Section~6.

In the quadratic order, the $00$-component of the  Ricci tensor $R_{00}$ is given by 
\begin{equation}
\label{00secondricci}
R^{(2)}_{00}=\frac{1}{2} \Delta h^{(2)}_{00} +2\Phi \Delta \Phi -2\Phi_{,i} \Phi_{,i} \; ,
\end{equation}
where the static ansatz is assumed. The last two terms on the r.h.s. are sourced by the linear metric perturbations. These, we remind, are the same as in GR and are given by Eq.~\eqref{metrictest}.  In the second order, the standard matter contributes to the 
r.h.s. of Einstein equations by
\begin{equation}
\label{contrst}
\left(T^{matter}_{00}-\frac{1}{2}T^{matter}g_{00} \right)^{(2)} = {\cal A} ({\bf r})  \Phi \; ,
\end{equation}
where ${\cal A} ({\bf r})$ is the mass distribution inside the Sun, $\int d{\bf r} {\cal A} ({\bf r}) =M$. Here we explicitly neglect the pressure of the Sun as well as the relative motion of its components. 
The analogous contribution due to IDM is inferred from Eq.~\eqref{mimstress},
\begin{equation}
\nonumber
\left(T^{IDM}_{00}-\frac{1}{2} T^{IDM} g_{00} \right)^{(2)} =\frac{1}{2}\Sigma^{(2)} \; .
\end{equation}
Note that we ignored the term $\sim (\square \varphi)^2$ present in Eq.~\eqref{mimstress}. According to Eq.~\eqref{lindeltasun}, it may be 
non-zero only in the non-perturbative region. To write the resulting expression for the second order metric perturbation $h^{(2)}_{00}$, we switch to the spheric coordinates, where 
it takes the elegant form, 
\begin{equation}
\label{002}
h^{(2)}_{00}= 2\Phi^2+\frac{8\pi \chi }{M^2_{Pl}} \cdot \frac{\Phi^2}{v^4} \cdot \frac{1}{(1-\cos \theta)^2}-4 \tilde{\Phi} \; ;
\end{equation}
$\theta$ is the angle between the $z$-axis and the line of sight. The potential $\tilde{\Phi}$ is defined by
\begin{equation}
\label{tildephi}
\tilde{\Phi}= -\frac{1}{M^2_{Pl}}\int \frac{{\cal A} ({\bf y}) \Phi ({\bf y})d{\bf y}}{|{\bf x}-{\bf y}|} \; .
\end{equation}
Notably, this part of the metric perturbation caused by `self-gravity' of the Sun carries no imprint of IDM. We relegate the details of derivation of Eq.~\eqref{002} to Appendix~D. 
Here let us note the main trick used at this level. Instead of integrating out directly Eq.~\eqref{sigma2}, it is more convenient to write the expression for the field $\Sigma^{(2)}$ as follows, 
\begin{equation}
\label{replace}
\Sigma^{(2)}= \frac{\chi}{v} \Delta \int^{z}_{-\infty} \Delta \delta \varphi^{(2)} \; .
\end{equation}
Then, the IDM induced metric correction can be obtained immediately 
\begin{equation}
\label{metriccormain}
\delta h^{(2)}_{00} =\frac{8\pi \chi}{M^2_{Pl}v} \int^{z}_{-\infty} \Delta  \delta  \varphi^{(2)} d\tilde{z}. 
\end{equation}
Using Eq.~\eqref{int} from Appendix~D, we reproduce the second term on the r.h.s. of Eq.~\eqref{002}.

In the limit $\theta \simeq \frac{\rho}{z} \rightarrow 0$, this term grows fast and quickly becomes of the 
order of the main terms in the metric expansion. For very small values of the angle $\theta$, however, 
the perturbative description employed in the present Section is not valid, and one may expect softening of the singularity in the non-perturbative phase. This is indeed the 
case, as we will see in Section~6.  On the other hand, the expression~\eqref{002} remains finite upstream from the Sun ($\theta \simeq \pi$), where it tends to the constant value (with respect to the angle $\theta$) $\delta h^{(2)}_{00} \simeq \frac{\chi }{M^2_{Pl}} \cdot \frac{\Phi^2}{v^4}$. Therefore, gravitational focusing has a marginal impact on the planetary dynamics, 
even if they never pass through the downstream region.

As for the concluding remark, we would like to stress that Eqs.~\eqref{metrictest} and~\eqref{002} represent only a particular solution of the Einstein equations (see the discussion 
in the end of Section~4). The general solution for the metric perturbation $h_{00}$ will be fixed in Section~6 upon defining the 
full IDM profile.

\section{Going beyond perturbative regime}

 To understand better the region of the applicability of the perturbative analysis, let us calculate the third order correction to the field $ \varphi$. Using the same 
arguments as in the beginning of Section~5, we write down the simplified version of the relevant equation, 
\begin{equation}
\nonumber 
-\partial_i \delta \varphi^{(3)} v^i + \partial_i \delta \varphi^{(1)} \partial_i \delta \varphi^{(2)} =0 \; .
\end{equation}
Once again, higher order corrections to the field $\varphi$ are defined by the lower order ones, while metric perturbations do not directly affect them. The solution 
for this equation is given by
\begin{equation}
\label{smallrhovelocity3}
\delta \varphi^{(3)} = \frac{8M^3}{v^5 M^6_{Pl} }  \frac{z^2}{\rho^4} \; ,
\end{equation}
where we restrict to the case of small variables $\rho$, still obeying the inequality~\eqref{cond}. 
Despite the strong amplification in the limit $\rho \rightarrow 0$, the hierarchy $|\delta \varphi^{(3)}| \ll  |\delta \varphi^{(2)}| \ll |\delta \varphi^{(1)}|$ holds everywhere in the region~\eqref{cond}. On the other hand, 
for $\frac{\rho}{z} \sim \sqrt{\frac{|\Phi|}{v^2}}$, one has $|\delta \varphi^{(1)}| \sim |\delta \varphi^{(2)}| \sim |\delta \varphi^{(3)}|$, what implies the breakdown of the perturbation approach. By recursion, 
that statement can be extended to an arbitrary order. We conclude that Eq.~\eqref{cond} remains intact upon including higher order corrections.

Now let us compare the leading order asymptotic behavior of the corrections 
${\bf  \nabla} \delta \varphi^{(1)}$, ${\bf \nabla} \delta \varphi^{(2)}$ and ${\bf \nabla} \delta \varphi^{(3)}$ in the region~\eqref{cond}. We obtain from Eqs.~\eqref{smallrhovelocity2} and~\eqref{smallrhovelocity3}, 
\begin{equation}
\nonumber 
{\bf \nabla} \delta \varphi^{(1)}, {\bf \nabla} \delta \varphi^{(2)}, {\bf \nabla} \delta \varphi^{(3)} \propto  \left (\frac{z}{\rho} {\bf e}_{\rho} -\frac{{\bf e}_z}{2}\right) \; .
\end{equation}
By recursion, one can check that the same holds to an arbitrary order, 
\begin{equation}
\nonumber
\nabla \delta \varphi^{(n)}=F_n \cdot \left(\frac{z}{\rho} {\bf e}_{\rho} -\frac{{\bf e}_z}{2}\right) \; .
\end{equation}
That is, in the limit $\rho \rightarrow 0$, corrections to the velocity ${\bf v}=-{\bf \nabla} \bar{\varphi}$ possess the same directional dependence being 
different only in the magnitude.

This simple observation allows to assess the relevant fields in the deeply non-linear phase, 
\begin{equation}
\label{condnonpert}
\frac{\rho^2}{z^2} \ll \frac{|\Phi|}{v^2} \; .
\end{equation}
The following ansatz for the perturbation ${\nabla} \delta \varphi$ should work,
\begin{equation}
\label{nablavarphiinter}
\nabla \delta \varphi = \nabla \delta \varphi^{(1)}+F \cdot \left(\frac{z}{\rho} {\bf e}_{\rho} -\frac{{\bf e}_z}{2}\right)  \; .
\end{equation}
For the sake of future convenience, we explicitly separated  the linear order term. In the perturbative phase, the function $F$ is defined as a series $F =\sum^{\infty}_{n=2} F_n$. 
Our task is to find the function $F$ suitable for both regimes~\eqref{cond} and~\eqref{condnonpert}. For this purpose, we substitute the ansatz~\eqref{nablavarphiinter}  into the equation 
\begin{equation}
\nonumber 
-2 \partial_i \delta \varphi v^i+\partial_i \delta \varphi \partial_i \delta \varphi =-2\Phi \; .
\end{equation}
We again make use of the Newtonian approximation for the metric. This gives for the function $F$,
\begin{equation}
\label{F}
F  = -\frac{2M}{vM^2_{Pl}} \frac{1}{z} -\frac{v}{2}\frac{\rho^2}{z^2}+\frac{\rho}{z}\sqrt{\frac{2M}{M^2_{Pl}z} +\frac{v^2}{4} \frac{\rho^2}{z^2}} \; ,
\end{equation}
where we omitted the terms of the order $\frac{\rho^2}{z^2} \times$(those present here).
Combining everything altogether, 
we obtain 
\begin{equation}
\label{velocityinter}
{\bf \nabla} \delta \varphi= -\left(\frac{v}{2} \frac{\rho^2}{z^2}-\frac{\rho}{z} \sqrt{\frac{2M}{M^2_{Pl}z}+\frac{v^2 \rho^2}{4 z^2}} \right) \left( \frac{z}{\rho}{\bf e}_{\rho}-\frac{{\bf e}_z}{2} \right) \; .
\end{equation}
Expanding this expression in powers of the potential $\Phi$, one reproduces our perturbative results, as it should be. Furthermore, it matches the velocity 
distribution, which characterizes a hypothetic flow of dust particles past the point source. See Appendix~B for the exact formulae. Here let us point out the cancellation between the term ${\bf \nabla} \delta \varphi^{(1)}$ in Eq.~\eqref{nablavarphiinter} and the first term on the r.h.s. of Eq.~\eqref{F}. Each of them taken separately would 
lead to the singular behavior of the resulting velocity perturbation in the limit $\rho \rightarrow 0$. Due to their cancellation, the expression~\eqref{velocityinter} remains finite. This is the smoothening 
of the field $\varphi$ (and, consequently, $\Sigma$) quoted in the previous Sections. 
Such a smoothening in the non-linear regime is similar---albeit different in nature---to the Vainshtein screening (see e.g. the review on the Vainshtein mechanism~\cite{Babichev:2013usa}).

The expression~\eqref{velocityinter} is enough in order to calculate higher derivatives of the field $\varphi$ and the energy density $\Sigma$ in the non-perturbative phase. In particular, the 
field $\Sigma$ is defined by the equation analogous to Eqs.~\eqref{sigma1} and~\eqref{eqdefsigma2}, but with one important modification. Since derivatives with respect to the variable $\rho$ are more relevant now,  one makes the replacement
\begin{equation}
\nonumber 
\partial_i \left(\Sigma v^i \right)  \rightarrow -\frac{1}{\rho} \partial_{\rho} \left( \rho \Sigma \partial_{\rho} \delta \varphi \right) \; .
\end{equation}
Using Eq.~\eqref{velocityinter} and performing integration, we find the leading order asymptotics in the limit $\rho \rightarrow 0$ for the density field $\Sigma$,
\begin{equation}
\label{sigmanonpert}
\Sigma = \frac{\chi}{\rho^2} +{\cal O} \left(\frac{1}{\rho} \right)\; .
\end{equation}
Note a milder dependence on the variable $\rho$ compared to Eq.~\eqref{sigmapertzplus}. Interestingly, the Lagrange multiplier field $\Sigma$ is independent of the mass of the source. 
This appears to be a generic feature of the static solutions in the highly non-linear regime, see Appendix~A. From Eq.~\eqref{sigmanonpert} it is straightforward to evaluate the correction 
to the Newtonian potential induced by IDM, 
\begin{equation}
\label{00nonpert}
\delta h_{00}= \frac{4\pi \chi}{M^2_{Pl}} \ln^2 \frac{\theta}{\theta_c} +C \; .
\end{equation}
Here $C$ and $\theta_c$ are the constants of integration. In particular, the constant $\theta_c$ plays the role of the cutoff on the angles $\theta$. 
In the limit $\theta_c  \rightarrow 0$, the solution~\eqref{00nonpert} blows up. This corresponds to the situation, when 
the energy density $\Sigma$ retains its non-integrable $\rho$-dependence as in Eq.~\eqref{sigmanonpert} arbitrarily close to the positive $z$-axis. 
One may expect that new physics or unaccounted phenomenon smoothen the singularity in Eq.~\eqref{sigmanonpert} leading to a finite cutoff $\theta_c$. 
We will concretize the possible values of $\theta_c$ shortly.

Beforehand, let us address one long-standing issue: how does the non-perturbative part of the IDM profile affect the metric in the region~\eqref{cond}? The full metric correction can be 
written as follows,
\begin{equation}
\label{00pertplus}
\delta h^{full}_{00} \simeq \delta h_{00} +h^{bound}_{00} \; .
\end{equation} 
Here $\delta h_{00}$ is the particular solution of the Einstein equation drawn from the perturbatively calculated IDM profile. We remind that it is zero in the linear order and is given by the second term of the r.h.s. Eq.~\eqref{002} in the quadratic order. 
The term $h^{bound}$ accounts for IDM distribution in the non-regime. It is estimated by 
\begin{equation}
\label{bound}
h^{bound}_{00} \simeq - \frac{8\pi \chi}{M^2_{Pl}} \ln \theta \ln \theta_c \; .
\end{equation}
Consistently, it satisfies the homogeneous Einstein equation $R_{00}=0$ reducing to the Laplace equation $\Delta h^{bound}_{00}=0$. The function $h^{bound}_{00}$ is set by matching between the full metric correction~\eqref{00pertplus} and the value~\eqref{00nonpert} 
on the boundary $\theta \sim \sqrt{|\Phi|/v^2}$.

We see that the term $h^{bound}_{00}$ stemming from the non-perturbative part of the IDM profile gives a large contribution to the overall metric correction. 
Namely, it is enhanced at least logarithmically relative to the particular solution $\delta h_{00}$. The two contributions, however, are clearly non-degenerate. First, 
non-perturbative effects are independent of the Sun's mass. Second, the terms $\delta h_{00}$ and $h^{bound}_{00}$  have different shapes with respect to the variables $\theta$ and $r$. 
This ensures that they can be treated separately, when probing our model with existing observational data.

For viability of the IDM scenario, we should assume that the growth of the energy density ~\eqref{sigmanonpert} is cut at some small $\rho$, what translates into 
a finite cutoff $\theta_c$. 
To probe possible values of $\theta_c$, let us first elucidate the origin of the singularity. 
Its appearance can be traced back to the $\rho$-component of the velocity perturbation ${\nabla} \delta \varphi$, which does not vanish in the limit 
$\rho \rightarrow 0$. Therefore, a caustic is formed nearby the positive $z$-axis. This result is not particularly surprising, and could be anticipated readily from the 
existence of the constraint~\eqref{constraint}. We see that the sharp growth of the energy density in the limit $\rho \rightarrow 0$ can 
be smoothened, once the singularity at the caustic is cured.

In the end of Section~3, it was pointed out that the caustic singularity may be resolved apart from the 
test field approximation. The contribution due to IDM is sub-dominant, if the (derivative of) metric correction~\eqref{00nonpert} 
is not larger than the (derivative of) Newtonian potential. This condition translates into an inequality for the angle $\theta$,
\begin{equation}
\label{testlimit}
\theta \gtrsim \frac{8\pi \chi}{M^2_{Pl}} \cdot \frac{|\ln \theta_c |}{|\Phi|} \; .
\end{equation}
We substitute the characteristic value of the potential $\Phi$ in the Solar system $|\Phi| \sim 10^{-8}$. 
As for the constant $\chi$, we borrow the upper bound from the next Section, $\chi/M^2_{Pl} \sim 10^{-18}$. 
Then the condition~\eqref{testlimit} gives $\theta \gtrsim 4 \cdot 10^{-8}$. Thus, if the caustic singularity 
is cured beyond the test field approximation, we get in terms of the logarithm of the angular cutoff $|\ln \theta_c| \gtrsim 20$.

Otherwise, one attributes smoothening of the IDM energy density to the UV completing operators. Recall that the necessity of the UV cutoff by itself is not a new 
assumption. Indeed, the model~\eqref{action} is plagued with the ghost instabilities, 
which lead to a rapid vacuum destabilization, unless there is a 
sufficiently low cutoff/strong coupling scale. The latter is constrained as $\Lambda \lesssim 10~\mbox{TeV}$. 
It is reasonable to assume that the caustic singularity is resolved at those or smaller scales. 
This implies the constraint on the logarithm of the angular cutoff  $|\ln \theta_c| \lesssim 70$. 

We summarize that the cutoff $\theta_c$ is bounded as 
\begin{equation}
\nonumber 
20 \lesssim |\ln \theta_c| \lesssim 70 \; .
\end{equation}
For those tiny values of $\theta_c$, the metric correction $h^{bound}$ sourced by distribution of IDM on the boundary is amplified at least by the factor $\sim 10 \div 100$ compared 
to $\delta h_{00}$.

\section{Solar system constraints}

Our purpose in the present Section is to constrain the theory parameter $\chi$ starting from the non-observation 
of deviations from GR in the Solar system. As we have seen in the Section 4, to the linear order, gravitational focusing experienced by the flow of IDM produces no effect 
on the metric perturbations. Thus, no constraints on the constant $\chi$ proceed at this level. 
On the contrary, vector/scalar-tensor models with the preferred frame, e.g., 
Einstein-Aether and closely related setups~\cite{Blas:2010hb, Foster:2005dk}, generically predict non-zero PPN parameters $\alpha_1$ and $\alpha_2$. Of course, preferred frame effects are not absent in the IDM scenario, but simply emerge in a different way. 
In particular, the structure of the energy density $\Sigma$ and metric corrections exhibit a clear directional ($\theta$-) dependence.

These second order 
corrections yield non-trivial constraints on the theory constant $\chi$. For the sake of convenience, we rewrite Eq.~\eqref{002} 
in such a form, that the deviation of the $00$-component of the metric from its GR counterpart $h^{GR}_{00}$ is manifest,
\begin{equation}
\label{deviation}
h_{00}-h^{GR}_{00}= 2(\beta -1) \Phi^2 \; .
\end{equation}
Here we introduced the parameter $\beta$ (not to be confused with the correct PPN parameter $\beta_{PPN}$) given by
\begin{equation}
\label{betaPPN}
\beta= 1+\frac{4\pi \chi}{M^2_{Pl}v^4} \cdot \frac{1}{(1-\cos \theta)^2} \; .
\end{equation}
The above defined quantity $\beta$ is similar to the standard PPN parameter $\beta_{PPN}$. Therefore, we expect the metric correction~\eqref{deviation} to influence the precession of the Mercury perihelion, the subject of high precision measurements nowadays. 
Non-observation of any deviations from GR at this level implies a stringent bound on the theory constant $\chi$  inferred from the 
existing limits on the PPN parameter $\beta_{PPN}$. We would like to stress, however, that the parameters $\beta$ and $\beta_{PPN}$ are not identical, 
as the former explicitly depends on the spatial coordinate $\theta$, while the latter is assumed to be constant. To be on the safe side, we formulate our goal as to {\it conservatively} bound the theory constant $\chi$. 
This can be done rigorously for the following reasons. First, the quantity $(\beta-1)$ is always positively defined. This ensures that the effect encoded in Eq.~\eqref{deviation} does not average 
to zero, when considered on large time scales. Furthermore, the quantity $(\beta-1)$ exhibits a monotonous growth towards the smaller angles $\theta$ 
starting from the minimal value\footnote{The actual minimal value for the quantity $\beta-1$ is the quarter of the r.h.s. of Eq.~\eqref{betasmallangle}. Our sloppy estimate, however, better fits to the realistic physical situation discussed below.}
\begin{equation}
\label{betasmallangle}
\beta -1\simeq \frac{4\pi \chi}{M^2_{Pl} v^4} \; ,
\end{equation}
which sets the lower bound on the deviation from GR in the IDM scenario. This 'minimal' deviation has manifestly a PPN form, 
and thus the bounds on the parameter $\beta_{PPN}$ can be used to constrain the constant $\chi$.

In fact, the choice $\theta \simeq 1$ implied in Eq.~\eqref{betasmallangle} appears to be the most reasonable one. Namely, it corresponds to the case, when the preferred and halo rest frames coincide. The latter moves relative to the Sun with the 
speed $v \approx 220~\mbox{km/s}$ in the direction inclined towards the ecliptic plane by the angle $\theta \simeq \frac{\pi}{3} $. Hence, the use of the estimate~\eqref{betasmallangle} is 
fully justified also from the viewpoint of the physically realistic conditions. According to the discussion above, we identify the parameter $\beta$ at its minimal value to the PPN parameter $\beta_{PPN}$. We then make use of the observational bounds
 deduced from the study of the precession of the Mercury perihelion~\cite{Fienga, Verma:2013ata}, 
\begin{equation}
\label{betaexperiments}
\beta_ {PPN}-1 =(-4.1 \pm 7.8) \times 10^{-5} \; .
\end{equation}
We arrive at the following limit for the constant $\chi$, 
\begin{equation}
\label{constraintbeta}
 \frac{\chi}{M^2_{Pl}} \lesssim  10^{-18} \; .
\end{equation} 
This constraint carries a certain advantage over the limits of Eqs.~\eqref{limitstructure} and~\eqref{micros}. 
In particular, our constraint is valid independently on whether IDM gives the main contribution to the overall DM (as it is assumed in Eq.~\eqref{limitstructure}), or it constitutes only a tiny fraction. Besides, it is not sensitive to details of microscopic physics.

Stronger constraints on the parameter $\chi$ would follow, if the velocity ${\bf v}$ was inclined towards the ecliptic plane by the small angle $\bar{\theta} \ll 1$, see Fig.~\ref{plot}. For example, this is the case, if the 
preferred frame is at rest with respect to the cosmic microwave background. Note that the Sun moves relative to the cosmic frame with the speed $v \approx 369~\mbox{km/s}$ in the direction characterized by the 
angle of inclination $\bar{\theta}\simeq \frac{\pi}{18}$~\cite{planck}. In that case, the effects of gravitational focusing are more pronounced. For simplicity, let us fix the angle $\theta$ entering~\eqref{betaPPN} at the value $\bar{\theta} = \frac{\pi}{18}$. 
This is by no means a realistic assumption, since the parameter $\beta$ undergoes the large modulation during the period of the Mercury orbiting in the situation of interest. We make it merely to probe the sensitivity 
of the Mercury perihelion observations to the theory constant $\chi$. The result reads
\begin{equation}
\label{conservative}
 \frac{\chi}{M^2_{Pl}} \lesssim  10^{-21} \; .  
\end{equation}
To be more specific, one should consider the peculiar motion of Mercury with respect to the direction of the velocity ${\bf v}$. This task 
is out of the scope of the paper.

The constraint~\eqref{conservative} is not trustworthy also for another reason. It is unlikely that the preferred and cosmic frames coincide. 
Indeed, IDM residing in the halo is decoupled from the cosmological background. Still, this discussion may be of some interest for the following considerations. 
The limit~\eqref{constraintbeta} assumes that IDM has a zero velocity everywhere in the halo. The real picture is somewhat more complicated, as the Galaxy may be permeated by the 
local flows~\cite{Sikivie:2002bj}, which sum up to give the zero net momentum of the overall DM distribution. This our ignorance about the halo physics in the IDM scenario leads to the uncertaintity, when 
we attempt to quantify an IDM flow past the Sun. While barring the fine-tuning, the velocity $v$ is estimated by $v \simeq 10^{-3}$, its inclination 
towards the ecliptic plane is less constrained. Therefore, there is a room left for strengthening the bound~\eqref{constraintbeta} on the parameter $\chi$ with the better understood halo physics.

As for concluding remarks, let us discuss our result~\eqref{constraintbeta} in relation to other constraints. 
The limit~\eqref{constraintbeta} is many orders of magnitude stronger than the one concluded from the structure 
formation considerations~\cite{Capela:2014xta} given in Eq.~\eqref{limitstructure}. As a result, once Eq.~\eqref{constraintbeta} is satisfied, IDM behaves identically to CDM at the cosmological scales. 
On the other hand, IDM leaves drastically different signatures at the level of the Solar system, compared to CDM. 
In particular, using the expression~\eqref{sigmapertzplus}, we can estimate the energy density of IDM. 
Taking $\frac{\chi}{M^2_{Pl}} \sim 10^{-18}$ and $\theta \simeq 1$, 
we obtain at the Earth distance 
\begin{equation}
\nonumber 
\Sigma \simeq 1.5 \times 10^5 ~\frac{\mbox{GeV}}{\mbox{cm}^3} \; .
\end{equation}
This is to be contrasted to the background energy density of DM near the Earth $\Sigma \sim 0.3~ \frac{\mbox{GeV}}{\mbox{cm}^3}$. Downstream from the Sun, 
where the effects of gravitational focusing are particularly prominent, the energy density 
is further enhanced by the factor $1/\theta^6$. These estimates justify neglecting the 'inevitable' part of IDM related to its presence in the halo.

Finally, we would like to elucidate some implications of the constraint~\eqref{constraintbeta} for the microscopic physics. 
The strong coupling in the model at hand is given by $\Lambda_{str} \sim \frac{\chi^{3/4}}{M^{1/2}_{Pl}}$~\cite{Koyama:2009hc, Blas:2010hb, Ramazanov:2016xhp}. 
Consequently, with the limit~\eqref{constraintbeta}, we expect the UV physics to come into play at the scales $\Lambda \lesssim \Lambda_{str} \lesssim 300$ TeV. 
Interestingly, within one or two orders of magnitude, this expectation meets the microscopic physics requirement~\cite{Ramazanov:2016xhp}: $\Lambda \lesssim 10$ TeV. 
Note that the strong coupling scale/the scale of UV physics cannot be arbitrarily small, $\Lambda_{str} \gtrsim \Lambda \gtrsim (0.1~\mbox{mm})^{-1}$---otherwise, 
one may run into the conflict with the sub-mm tests of the Newton's law. This translates into the lower bound on the theory constant $\chi/M^2_{Pl} \gtrsim 10^{-42}$. 
Given our upper bound~\eqref{constraintbeta}, we conclude that there is still a large window for the parameter $\chi$ worth to explore.

We showed that the Solar system is a useful playground for testing IDM. We leave for future studies of astrophysical systems with compact objects, 
e.g., pulsars and black holes. These are believed to provide alternative (hopefully, 
even more stringent) limits on the parameter $\chi$. 

\vspace{0.5cm}
{\bf Acknowledgments.} We would like to thank Yuri Eroshenko for helpful correspondence on Cold Dark Matter and Gilles~Esposito-Far\`ese for 
discussions on the Solar system tests and critical remarks. 
E.B. was supported in part by the research program, ``Programme national de cosmologie et galaxies'' of the CNRS/INSU, France, 
and Russian Foundation for Basic Research Grant No. RFBR 15-02-05038.

\section*{Appendix A: Spherically symmetric solutions}

The present Appendix serves to assess the modification of gravity produced in the regime, where the inequality~\eqref{inequality} is not obeyed. 
For the sake of simplicity, we will consider the limit of vanishing speed $v \rightarrow 0$. 
In this limit, the model possesses the spherical symmetry. A similar setup has been employed for the study of spherically symmetric objects in Horndeski/Galileon theory~\cite{Babichev:2011iz,Babichev:2012re,Babichev:2016jom,Cisterna:2015yla,Babichev:2016rlq, Cisterna:2016vdx, Maselli:2016gxk,Saito:2015fza}. 
There is an important difference, however: in the case of the Galileon scenario, the time-dependent ansatz is caused by the requirement 
that the local solution matches the cosmologically evolving scalar field. There are also scenarios where the Galileon does not evolve with time.
On the contrary, the time-dependence of the field $\varphi$ follows necessarily from the structure of the model~\eqref{action}.

The constraint equation~\eqref{constraint} then takes the form
\begin{equation}
\label{constrstatspher}
(\partial_r \delta \varphi)^2 =-2 \Phi \; .
\end{equation}
This gives for the field $\delta \varphi (r)$, 
\begin{equation}
\label{deltavarphi}
\delta \varphi (r) =\frac{2\sqrt{2Mr}}{M_{Pl}} \; , 
\end{equation}
modulo the irrelevant constant of integration.

In the spherically symmetric static case, the l.h.s. and r.h.s. of the (non)conservation equation~\eqref{conservation} can be written as
\begin{equation}
\nonumber 
\nabla_{\mu} \left(\Sigma \nabla^{\mu} \varphi \right) \rightarrow -\frac{1}{r^2} \frac{\partial}{\partial r} (r^2 \cdot \Sigma \partial_r \varphi) 
\end{equation}
and 
\begin{equation}
\nonumber
\square^2 \varphi \rightarrow \Delta^2 \varphi \; ,
\end{equation}
respectively; $\Delta =\frac{1}{r^2} \frac{d}{dr} r^2 \frac{d}{dr}$. Combining altogether, we get for the energy density $\Sigma$, 
\begin{equation}
\label{sigmastatspher}
\Sigma =-\chi \cdot \frac{\int r^2  \Delta^2 \delta \varphi (r) dr}{r^2 \partial_{r} \delta \varphi (r)}  \; .
\end{equation}
Performing the integration with respect to the variable $r$, we obtain
\begin{equation}
\label{Sigma}
\Sigma (r) =\frac{9}{4} \cdot \frac{\chi}{r^2} +\frac{C}{r^{3/2}}\; .
\end{equation}
Here $C$ is some constant of integration. It cannot be defined within the present analysis, as we are lacking the knowledge of IDM distribution inside the Sun. 
In the Newton's limit, only the $00$-component of the IDM stress energy tensor is relevant,
\begin{equation}
\nonumber 
T^{IDM}_{00}= \Sigma (r) \; .
\end{equation}
Here we omitted the $\chi$-terms in the expression for the IDM stress-energy tensor~\eqref{mimstress}: they are suppressed by the value of the 
potential $\Phi$.  The only relevant part of the Ricci tensor is given by $R_{00}=\frac{1}{2}\Delta h_{00}$. Consequently, Einstein equations reduce to 
\begin{equation}
\nonumber
\frac{1}{2}\Delta h_{00}= 4\pi G (M \delta ({\bf r}) +\Sigma (r)) \; . 
\end{equation}
The first term on the r.h.s. results into the standard Newton's law, while the second one encodes the correction to the gravitational potential,
\begin{equation}
\label{potentialsphere}
\delta h_{00} =\frac{18\pi \chi }{M^2_{Pl}} \ln\frac{ r}{r_0} +\frac{24  \pi }{M^2_{Pl}} C \sqrt{r} -\frac{2C'}{M^2_{Pl}r} \; ,
\end{equation}
where $r_0$ and $C'$ are the integration constant. 

As it follows, the correction to the Newtonian potential $\Phi \sim 1/r$ grows as $\sim r \ln r$ (relative to the potential $\Phi$). The different picture occurs in the regime $v^2 \gg |\Phi|$---in that case, IDM induced metric corrections 
do not exhibit any amplification with the radius $r$ (relative to the background terms in the metric expansion). See Eq.~\eqref{002}. To paraphrase, upon switching from the regime $v^2 \ll |\Phi|$ to $v^2 \gg |\Phi|$, 
the effects due to gravitational focusing become substantially milder. The same conclusion only strengthens, once the second term on the r.h.s. of Eq.~\eqref{potentialsphere} is included. 

Let us discuss the consequences, which the solution~\eqref{potentialsphere} entails for the results of the present work. As it has been pointed above, 
the value of the constant $C$ is defined upon matching of Eq.~\eqref{potentialsphere} to the solution for the metric inside the Sun. 
This task is out of the scope of this work and we do not consider the corresponding term in what follows. On the boundary $v^2 \sim |\Phi|$ the logarithmic 
term matches well the metric correction obtained from the perturbative distribution of IDM, i.e., the second term on the r.h.s. of Eq.~\eqref{002}. On the other hand, 
the term $\sim 1/r$ tells us something new. In the certain sense, it is analogous to the  'boundary' term~\eqref{bound} and stands for the 
discrepancy between the particular solution of Einstein equations~\eqref{metrictest} and~\eqref{002} from the full solution. 
This discrepancy is now sourced by the distribution of IDM in the region~$v^2 \lesssim |\Phi|$, with the constant $C'$ playing the 
role of the mass of IDM in that region. Namely, $C' \simeq \int_{v^2 \lesssim |\Phi|} \Sigma ({\bf r}) d{\bf r} \sim \chi \cdot D$, where $D$ denotes 
a characteristic size of the region. 

Therefore, the term $\sim 1/r$ should be included in the full metric solution~\eqref{002}. Let us 
argue, however, that it does not lead to any physically observable consequences, and, hence, 
does not compromise the results of the present research. Indeed, according to its $r$-dependence, 
it can be eliminated by simultaneously shifting the gravitational constant compared to the Newton's constant\footnote{Here it is relevant that the same term $\sim \frac{C'}{r}$ also contributes to the $ij$-component of the metric. 
Hence, shifting the gravitational constant, it is eliminated at this level as well.} $G_{N} =\frac{1}{M^2_{Pl}}$. The shift is estimated by $\frac{G}{G_N}-1 \sim \frac{\chi \cdot D}{M} \sim \frac{\chi}{M^2_{Pl} |\Phi|}$. This is unobservable 
given our stringent limit on the constant $\chi$.

\section*{Appendix B: Flow of dust particles past the point source}
In Section~2, it was pointed out that Eq.~\eqref{constraint} leads to the 
geodesics equation followed by test particles in the external gravitational field. We also observed in the 
main part of the paper, that the Newtonian limit works fairly well, when evaluating the corrections to the field 
$\varphi$. This suggests an alternative way to calculate the velocity potential $\varphi$, or, more precisely, 
its spatial derivatives. Namely, one tracks the flow of dust particles in the gravitational field created by the point source. Hereafter, we follow Ref.~\cite{Sikivie:2002bj}, 
and parametrize the trajectory of a single particle as
\begin{equation}
\nonumber
t=\frac{a}{v} \left(e \cdot \mbox{sinh} \Psi-\Psi \right) \; ,
\end{equation}
\begin{equation}
\label{zapp}
z(b, \Psi) =a \left(e \cdot \mbox{sinh} \Psi +1-\frac{1}{e} \mbox{exp}  \Psi \right)
\end{equation}
and 
\begin{equation}
\label{rhoapp}
\rho (b, \Psi) =b \left(1-\frac{1}{e}\mbox{exp} \Psi \right) \; .
\end{equation}
Here $b$ is the impact parameter, $\Psi$ is the eccentric anomaly parameter, $e=\sqrt{1+\frac{b^2}{a^2}}$, and $a =\frac{GM}{v^2}$. 
In terms of the parameters $\Psi, b, e$ and $a$, the velocity of the flow is described by
\begin{equation}
\label{vrho} 
v_{\rho}=-v\frac{b\cdot \mbox{exp} \Psi}{e \cdot a \left(e\cdot \mbox{cosh} \Psi-1 \right)} \; ,
\end{equation}
\begin{equation}
\label{vz}
v_z =v \frac{e \cdot \mbox{cosh} \Psi -\frac{1}{e} \cdot \mbox{exp} \Psi}{e\cdot \mbox{cosh} \Psi-1} \; .
\end{equation}
One then inverts Eqs.~\eqref{zapp} and~\eqref{rhoapp}, so that to express the parameters $b$ and $\Psi$ via the variables $z$ and $\rho$, 
\begin{equation}
\label{b}
b_{\pm}=\frac{\rho}{z} \left(1\pm \sqrt{1+y} \right) 
\end{equation}
and
\begin{equation}
\label{psi}
\Psi_{\pm}=\ln \left(e_{\pm} \cdot \frac{1}{y} \left[ 1 \mp \sqrt{1+y}\right]^2 \right) \; ,
\end{equation}
where $y$ and $e_{\pm}$ are given by 
\begin{equation}
\label{y}
y=\frac{4a\cdot  (z+\sqrt{\rho^2+z^2})}{\rho^2} 
\end{equation}
and 
\begin{equation}
\nonumber 
e_{\pm} =\sqrt{1+\frac{b^2_{\pm}}{a^2}} \; ,
\end{equation}
respectively. Substituting Eqs.~\eqref{b} and~\eqref{psi} into Eqs.~\eqref{vrho} and~\eqref{vz}, one finally obtains the velocity of the flow as the function of the spatial variables $z$ and $\rho$. 
This should be contrasted to the results obtained in the bulk of the paper. There is a subtlety. In the case of the particle DM, two solutions labeled by the subscripts $'+'$ and $'-'$ correspond to the two flows produced 
at the shell-crossing. On the other hand, IDM is fundamentally single flow, as it has been pointed out in the end 
of Section~3. Therefore, out of two, we should pick one solution. The solution must satisfy our boundary conditions for the field $\varphi$: ${\bf \nabla} \delta \varphi \rightarrow 0$ at $z \rightarrow -\infty$. 
Only the branch with the upper signs in the equations above fulfills this condition. We stick to this branch in what follows. 
 For simplicity, we consider only 
the radial component of the velocity $v_{\rho}$. This is given by
\begin{equation}
\label{veldustflow}
v_{\rho}=\frac{4va}{\rho \cdot y \cdot {\cal F} (\rho, y)} \cdot \left(1-\sqrt{1+y} \right) \; ,
\end{equation}
where 
\begin{equation}
\nonumber
{\cal F} (\rho,y) =1+\frac{4a^2}{\rho^2} \cdot \frac{[1-\frac{1}{y} (1-\sqrt{1+y})^2]^2}{[1-\sqrt{1+y}]^2} \; .
\end{equation}
In the limit $\rho \rightarrow 0$ and for positive $z$, the argument $y$ blows up, i.e., $y\rightarrow \infty$. Simultaneously, ${\cal F} (\rho, y) \rightarrow 1$, and one arrives at
\begin{equation}
\nonumber 
v_{\rho} \rightarrow -\sqrt{\frac{2M}{M^2_{Pl} z}} \; .
\end{equation}
This reproduces our result~\eqref{velocityinter}, where one should consider the limit $\rho \rightarrow 0$. At the same time, in the particle picture the perturbative regime corresponds to very small values of the variable 
$y$, i.e., $y \ll 1$. Expanding Eq.~\eqref{veldustflow} in a series over the powers of the parameter $y$, we get in the linear and in the quadratic order
\begin{equation}
\nonumber 
v^{(1)}_{\rho}=-\frac{2av}{\rho}
\end{equation} 
and 
\begin{equation}
\nonumber 
v^{(2)}_{\rho}=\frac{v\cdot a \cdot y}{4\rho} \left(  1+\frac{z}{\sqrt{\rho^2+z^2}}\right) \; ,
\end{equation}
respectively. These perfectly match Eqs.~\eqref{sollinear} and~\eqref{varphi2gen} from the 
bulk of the paper. 

\section*{Appendix C: Linear level analysis for the light source}

In Section~4, we concluded that IDM leaves no trace on the metric to the linear order. 
The discussion there was provided for a sufficiently compact object,---the Sun. In that case, 
the conditions~\eqref{inequality} and~\eqref{cond} break down near its surface and in the downstream region, respectively. 
To overcome this obstacle, we loosely extrapolated our perturbative calculation for the IDM fields 
$\Sigma$ and $\varphi$ to the whole space. One may inquire now: what is the structure of linear corrections to the metrics would be these inequalities trustworthy everywhere? That question is not of the pure heuristic interest. 
Indeed, for a sufficiently light and not very compact object like the Earth, the condition~\eqref{inequality} is obeyed 
down to its center. On the other hand, violation of the inequality~\eqref{cond} occurs only in a very narrow part of space 
measured by the angle $\theta \sim 10^{-2}$ near the surface of the Earth. We consider it a reasonable approximation 
to neglect possible effects, which stem from this region. Remarkably, despite the seeming differences between the cases of the Sun and the Earth, 
linear corrections to the metric are absent in the latter case either. This gives a further support for the main statement of 
Section~4: the effects due to gravitational focusing are completely irrelevant at the level of the linear analysis.

As for the first step, we find the field $\delta \varphi^{(1)}$, or, more precisely, $\Delta \varphi^{(1)}$. We apply the 
Laplace operator to both parts of Eq.~\eqref{lin}, and then integrate it out with respect to the variable $z$,
\begin{equation}
\label{linDeltareal}
\Delta \delta \varphi^{(1)}=\frac{4\pi}{M^2_{Pl}} \cdot \frac{1+2v^2}{v} \int^z_{-\infty} {\cal A} d\tilde{z} \; ,
\end{equation}
where ${\cal A} ({\bf r})$ denotes the mass distribution inside the Earth. In the linear order, the equation defining the 
field $\Sigma$ is given by Eq.~\eqref{sigma1}. Performing the integration with respect to the variable $z$ there and using Eq.~\eqref{linDeltareal}, we obtain 
\begin{equation}
\label{solsigmalin}
\Sigma^{(1)} =   \frac{4\pi (1+2v^2) \chi}{M^2_{Pl} \cdot v^2} \Delta \int^{z}_{-\infty}  dz' \int^{z'}_{-\infty} {\cal A} dz''\; . 
\end{equation}
We see that the energy density $\Sigma$ is zero everywhere, except for the tube-like region close to the positive $z$-axis. Away from it, the field $\Sigma^{(1)}$ does not 
affect metric perturbations, as it clearly follows from the structure of Eq.~\eqref{solsigmalin}. Hence, we can safely ignore it. 

It is straightforward to show that the $00$-metric component 
remains intact in the presence of IDM. Indeed, the latter contributes to the r.h.s. of Einstein equations by 
\begin{equation}
\nonumber 
\left(T^{IDM}_{00} -\frac{1}{2} T^{IDM} g_{00}\right)^{(1)}= \Sigma^{(1)} \left(\frac{1}{2}+v^2 \right) \; . 
\end{equation}
We set the latter to zero according to the discussion above. Hence, the standard matter is the only 
source of metric perturbations at this level. 

We then evaluate the $ij$-components of the metric. The relevant components of the Ricci tensor are given by
\begin{equation}
\nonumber 
R^{(1)}_{ij}=\frac{1}{2} \Delta h^{(1)}_{ij} \; ,
\end{equation}
where we made use of the harmonic gauge. 
The standard matter and IDM contribute to the r.h.s. of Einstein equations by
\begin{equation}
\nonumber 
\left(T^{matter}_{ij} -\frac{1}{2} T^{matter} g_{ij}\right)^{(1)}= \frac{1}{2} {\cal A}({\bf r}) \delta_{ij}  
\end{equation}
and
\begin{equation}
\nonumber 
\left( T^{IDM}_{ij}-\frac{1}{2} T^{IDM}g_{ij} \right)^{(1)}= \frac{1}{2}\Sigma^{(1)} \left( \delta_{ij} +v^i v^j\right) -\chi v^i \partial_j \Delta \delta \varphi^{(1)}-\chi v^j \partial_i \Delta \delta \varphi^{(1)} \; ,
\end{equation}
respectively. Neglecting the contribution from the field $\Sigma$, we obtain for the $ij$-metric components, 
\begin{equation}
\nonumber 
h^{(1)}_{ij}=2\Phi \delta_{ij} -\frac{16\pi \chi}{M^2_{Pl}} \left( v^i \partial_j \delta \varphi^{(1)}+v^j \partial_i \delta \varphi^{(1)}  \right) \; . 
\end{equation}
Finally, we calculate the $0i$-metric components. In the harmonic gauge, the relevant part of the Ricci tensor is given by
\begin{equation}
\nonumber 
R^{(1)}_{0i}=\frac{1}{2}\Delta h^{(1)}_{0i} \; .
\end{equation}
IDM contributes to the r.h.s. of Einstein equations by
\begin{equation}
\nonumber 
\left( T^{IDM}_{0i}-\frac{1}{2}T^{IDM}_{0i} \right)^{(1)} =\Sigma^{(1)} \partial_0 \varphi  v^i +\chi \partial_0 \varphi \partial_i \Delta  \delta \varphi^{(1)} \; ,
\end{equation} 
while the standard matter does not contribute at this level. Combining altogether, we obtain
\begin{equation}
\nonumber 
h^{(1)}_{0i}=\frac{16\pi \chi}{M^2_{Pl}}\partial_i \delta \varphi^{(1)}\; .
\end{equation}
Now let us perform the additional gauge transformation with 
\begin{equation}
\nonumber
\xi_{\mu}=\frac{16\pi \chi}{M^2_{Pl}} \partial_{\mu} \varphi \cdot  \delta \varphi^{(1)} \; .
\end{equation}
This serves to eliminate the anisotropic part in the metric components $h^{(1)}_{ij}$. 
Simultaneously, the components $h^{(1)}_{0i}$ turn into 0, while the $00$-component remains intact. We conclude that the linear order metric perturbations 
are unaffected by IDM also in the case of sufficiently light astrophysical objects. 

\section*{Appendix D. Derivation of Eq.~\eqref{002}} 
In the present Appendix, we discuss the details of derivation of Eq.~\eqref{002}. 
First, it is convenient to rewrite the expression for the second order of the 
$00$-component of the Ricci tensor~\eqref{00secondricci} as follows, 
\begin{equation}
\nonumber 
R^{(2)}_{00} = \Delta \left(\frac{1}{2}h^{(2)}_{00} +4 \tilde{\Phi} -\Phi^2 \right) \; ,
\end{equation}
where the potential $\tilde{\Phi}$ is defined by Eq.~\eqref{tildephi}. We rewrite the r.h.s. of Eq.~\eqref{contrst} in the similar fashion
\begin{equation}
\nonumber
{\cal A} ({\bf x}) \cdot \Phi ({\bf x}) =-\frac{1}{4\pi} \Delta \int \frac{{\cal A} ({\bf y}) \cdot \Phi ({\bf y})}{|{\bf x}-{\bf y}|} d{\bf y} \equiv \frac{1}{4\pi G} \Delta \tilde{\Phi}  \; .
\end{equation}
 Here we made use of the identity 
\begin{equation}
\nonumber
\Delta \frac{1}{|{\bf x}-{\bf y}|} =-4\pi \delta ({\bf x}-{\bf y}) \; .
\end{equation}
Finally, we express the second order energy density $\Sigma^{(2)}$ from Eq.~\eqref{eqdefsigma2}
\begin{equation}
\nonumber 
\Sigma^{(2)}= \frac{\chi}{v}\Delta \int^{z}_{-\infty} \Delta \delta \varphi^{(2)} d{\tilde{z}} \; .
\end{equation}
One makes use of Eq.~\eqref{varphi2gen} to evaluate the integral on the r.h.s. here, 
\begin{equation}
\label{int}
\int^z_{-\infty} \Delta \delta \varphi^{(2)}=\frac{\Phi^2}{v^3} \cdot \frac{1}{(1-\cos \theta)^2} \; .
\end{equation}
Combining altogether, we write the Einstein equation defining the field $h^{(2)}_{00}$
\begin{equation}
\label{simple}
\Delta \left(\frac{1}{2} h^{(2)}_{00} -F \right) = 0 \; ,
\end{equation}
where $F$ is the function given by 
\begin{equation}
\nonumber
F=\Phi^2-2\tilde{\Phi} +\frac{4\pi G}{v} \int^{z}_{-\infty} \Delta \delta \varphi^{(2)} d\tilde{z} =0 \; .
\end{equation}
Eq.~\eqref{simple} can be integrated out in the straightforward manner. The result is given 
by Eq.~\eqref{002} from the bulk of the paper.


\begin{thebibliography}{99}

\bibitem{Willbook}
C.~M.~Will, {\it Theory and Experiment in Gravitational Physics}, Cambridge University Press, Cambridge; New York (1993), 2nd edition. [Google Books].


\bibitem{Will:2014kxa}
  C.~M.~Will,
  Living Rev.\ Rel.\  {\bf 17} (2014) 4;
  [arXiv:1403.7377 [gr-qc]].

\bibitem{Chamseddine:2014vna}
  A.~H.~Chamseddine, V.~Mukhanov and A.~Vikman,
  JCAP {\bf 1406} (2014) 017;
  [arXiv:1403.3961 [astro-ph.CO]].

\bibitem{Capela:2014xta}
  F.~Capela and S.~Ramazanov,
  JCAP {\bf 1504} (2015) 051;
  [arXiv:1412.2051 [astro-ph.CO]].

\bibitem{Mirzagholi:2014ifa}
  L.~Mirzagholi and A.~Vikman,
  JCAP {\bf 1506} (2015) 028;
  [arXiv:1412.7136 [gr-qc]].

\bibitem{Jacobson:2000xp}
  T.~Jacobson and D.~Mattingly,
  Phys.\ Rev.\ D {\bf 64} (2001) 024028;
  [gr-qc/0007031].


\bibitem{Haghani:2014ita}
  Z.~Haghani, T.~Harko, H.~R.~Sepangi and S.~Shahidi;
  arXiv:1404.7689 [gr-qc].

\bibitem{Ramazanov:2015pha}
  S.~Ramazanov,
  JCAP {\bf 12} (2015) 007;
  [arXiv:1507.00291 [gr-qc]].

\bibitem{Horava:2009uw}
  P.~Horava,
  Phys.\ Rev.\ D {\bf 79} (2009) 084008;
  [arXiv:0901.3775 [hep-th]].

\bibitem{Sotiriou:2009gy}
  T.~P.~Sotiriou, M.~Visser and S.~Weinfurtner,
  Phys.\ Rev.\ Lett.\  {\bf 102} (2009) 251601;
  [arXiv:0904.4464 [hep-th]].

\bibitem{Sotiriou:2009bx}
  T.~P.~Sotiriou, M.~Visser and S.~Weinfurtner,
  JHEP {\bf 0910} (2009) 033;
  [arXiv:0905.2798 [hep-th]].

\bibitem{Koyama:2009hc}
  K.~Koyama and F.~Arroja,
  JHEP {\bf 1003} (2010) 061;
  [arXiv:0910.1998 [hep-th]].

\bibitem{Blas:2010hb}
  D.~Blas, O.~Pujolas and S.~Sibiryakov,
  JHEP {\bf 1104} (2011) 018;
  [arXiv:1007.3503 [hep-th]].

\bibitem{Ramazanov:2016xhp}
  S.~Ramazanov, F.~Arroja, M.~Celoria, S.~Matarrese and L.~Pilo,
  JHEP {\bf 1606} (2016) 020;
  [arXiv:1601.05405 [hep-th]].

\bibitem{Cline:2003gs}
  J.~M.~Cline, S.~Jeon and G.~D.~Moore,
  Phys.\ Rev.\ D {\bf 70} (2004) 043543;
  [hep-ph/0311312].

  \bibitem{Rubakov:2008nh}
  V.~A.~Rubakov and P.~G.~Tinyakov,
  Phys.\ Usp.\  {\bf 51} (2008) 759;
  [arXiv:0802.4379 [hep-th]].

\bibitem{Sreekumar:1997un}
  P.~Sreekumar {\it et al.} [EGRET Collaboration],
  Astrophys.\ J.\  {\bf 494} (1998) 523;
  [astro-ph/9709257].



\bibitem{Barvinsky:2015kil}
  A.~O.~Barvinsky, D.~Blas, M.~Herrero-Valea, S.~M.~Sibiryakov and C.~F.~Steinwachs,
  Phys.\ Rev.\ D {\bf 93} (2016) no.6,  064022;
  [arXiv:1512.02250 [hep-th]].

\bibitem{Mukohyama:2009mz}
  S.~Mukohyama,
  Phys.\ Rev.\ D {\bf 80} (2009) 064005;
  [arXiv:0905.3563 [hep-th]].




\bibitem{Blas:2009yd}
  D.~Blas, O.~Pujolas and S.~Sibiryakov,
  JHEP {\bf 0910} (2009) 029;
  [arXiv:0906.3046 [hep-th]].


\bibitem{Charmousis:2009tc}
  C.~Charmousis, G.~Niz, A.~Padilla and P.~M.~Saffin,
  JHEP {\bf 0908} (2009) 070;
  [arXiv:0905.2579 [hep-th]].

\bibitem{Blas:2009qj}
  D.~Blas, O.~Pujolas and S.~Sibiryakov,
  Phys.\ Rev.\ Lett.\  {\bf 104} (2010) 181302;
  [arXiv:0909.3525 [hep-th]].

\bibitem{Horava:2010zj}
  P.~Horava and C.~M.~Melby-Thompson,
  Phys.\ Rev.\ D {\bf 82} (2010) 064027;
  [arXiv:1007.2410 [hep-th]].

\bibitem{Chamseddine:2013kea}
  A.~H.~Chamseddine and V.~Mukhanov,
  JHEP {\bf 1311} (2013) 135;
  [arXiv:1308.5410 [astro-ph.CO]].

\bibitem{Deruelle:2014zza}
  N.~Deruelle and J.~Rua,
  JCAP {\bf 1409} (2014) 002;
  [arXiv:1407.0825 [gr-qc]].

\bibitem{Golovnev:2013jxa}
  A.~Golovnev,
  Phys.\ Lett.\ B {\bf 728} (2014) 39;
  [arXiv:1310.2790 [gr-qc]].


\bibitem{Hammer:2015pcx}
  K.~Hammer and A.~Vikman;
  arXiv:1512.09118 [gr-qc].

\bibitem{Arroja:2015wpa}
  F.~Arroja, N.~Bartolo, P.~Karmakar and S.~Matarrese,
  JCAP {\bf 1509} (2015) 051;
  [arXiv:1506.08575 [gr-qc]].

\bibitem{Arroja:2015yvd}
  F.~Arroja, N.~Bartolo, P.~Karmakar and S.~Matarrese,
  JCAP {\bf 1604} (2016) no.04,  042;
  [arXiv:1512.09374 [gr-qc]].

\bibitem{Danby} 
J.~Danby and G.~Camm, MNRS {\bf 117}, 50 (1957). 

\bibitem{Danby1}
J.~Danby and T.~Bray, Astron. J. {\bf 72}, 219 (1967).

\bibitem{Griest:1987vc}
  K.~Griest,
  Phys.\ Rev.\ D {\bf 37} (1988) 2703.

\bibitem{Sikivie:2002bj}
  P.~Sikivie and S.~Wick,
  Phys.\ Rev.\ D {\bf 66} (2002) 023504;
  [astro-ph/0203448].

\bibitem{Belotsky1}
  K.~M.~Belotsky, T.~Damour and M.~Y.~Khlopov,
  Phys.\ Lett.\ B {\bf 529} (2002) 10;
  [astro-ph/0201314].

\bibitem{Belotsky2}
  K.~Belotsky and M.~Khlopov,
  Grav.\ Cosmol.\  {\bf 11} (2005) 220
  [astro-ph/0504216].

\bibitem{Lee:2013wza}
  S.~K.~Lee, M.~Lisanti, A.~H.~G.~Peter and B.~R.~Safdi,
  Phys.\ Rev.\ Lett.\  {\bf 112} (2014) no.1,  011301;
  [arXiv:1308.1953 [astro-ph.CO]].


\bibitem{Patla:2013vza}
  B.~R.~Patla, R.~J.~Nemiroff, D.~H.~H.~Hoffmann and K.~Zioutas,
  Astrophys.\ J.\  {\bf 780} (2014) 158;
  [arXiv:1305.2454 [astro-ph.EP]].









\bibitem{Dubovsky:2004qe}
  S.~L.~Dubovsky,
  JCAP {\bf 0407} (2004) 009;
  [hep-ph/0403308].


\bibitem{Foster:2005dk}
  B.~Z.~Foster and T.~Jacobson,
  Phys.\ Rev.\ D {\bf 73} (2006) 064015;
  [gr-qc/0509083].

\bibitem{Blas:2014aca}
  D.~Blas and E.~Lim,
  Int.\ J.\ Mod.\ Phys.\ D {\bf 23} (2015) 13,  1443009;
  [arXiv:1412.4828 [gr-qc]].

\bibitem{planck}
  P.~A.~R.~Ade {\it et al.} [Planck Collaboration],
  Astron.\ Astrophys.\  {\bf 594} (2016) A5;
  [arXiv:1505.08022 [astro-ph.IM]].


\bibitem{Mukohyama:2009tp}
  S.~Mukohyama,
  JCAP {\bf 0909} (2009) 005;
  [arXiv:0906.5069 [hep-th]].










\bibitem{Lim:2010yk}
  E.~A.~Lim, I.~Sawicki and A.~Vikman,
  JCAP {\bf 1005} (2010) 012;
  [arXiv:1003.5751 [astro-ph.CO]].

\bibitem{Babichev:2016hys}
  E.~Babichev,
  JHEP {\bf 1604} (2016) 129;
  [arXiv:1602.00735 [hep-th]].



\bibitem{Izumi:2009ry}
  K.~Izumi and S.~Mukohyama,
  Phys.\ Rev.\ D {\bf 81} (2010) 044008;
  [arXiv:0911.1814 [hep-th]].

\bibitem{Greenwald:2009kp}
  J.~Greenwald, A.~Papazoglou and A.~Wang,
  Phys.\ Rev.\ D {\bf 81} (2010) 084046;
  [arXiv:0912.0011 [hep-th]].


\bibitem{Babichev:2013usa}
  E.~Babichev and C.~Deffayet,
  Class.\ Quant.\ Grav.\  {\bf 30} (2013) 184001;
  [arXiv:1304.7240 [gr-qc]].



\bibitem{Fienga}
  A.~Fienga, J.~Laskar, P.~Kuchynka, H.~Manche, G.~Desvignes, M.~Gastineau, I.~Cognard and G.~Theureau,
  Celest.\ Mech.\ Dyn.\ Astron.\  {\bf 111} (2011) 363;
  [arXiv:1108.5546 [astro-ph.EP]].


\bibitem{Verma:2013ata}
  A.~Verma, A.~Fienga, J.~Laskar, H.~Manche and M.~Gastineau,
  Astron.\ Astrophys.\  {\bf 561} (2014) A115;
  [arXiv:1306.5569 [astro-ph.EP]].















\bibitem{Babichev:2011iz}
  E.~Babichev, C.~Deffayet and G.~Esposito-Far\`ese,
  Phys.\ Rev.\ Lett.\  {\bf 107} (2011) 251102;
  [arXiv:1107.1569 [gr-qc]].

\bibitem{Babichev:2012re}
  E.~Babichev and G.~Esposito-Far\`ese,
  Phys.\ Rev.\ D {\bf 87} (2013) 044032;
  [arXiv:1212.1394 [gr-qc]].

\bibitem{Cisterna:2015yla} 
  A.~Cisterna, T.~Delsate and M.~Rinaldi,
  Phys.\ Rev.\ D {\bf 92}, no. 4, 044050 (2015);
  [arXiv:1504.05189 [gr-qc]].

\bibitem{Babichev:2016rlq}
  E.~Babichev, C.~Charmousis and A.~Leh\' ebel,
  Class.\ Quant.\ Grav.\  {\bf 33} (2016) no.15,  154002;
  [arXiv:1604.06402 [gr-qc]].

\bibitem{Cisterna:2016vdx}
  A.~Cisterna, T.~Delsate, L.~Ducobu and M.~Rinaldi,
  Phys.\ Rev.\ D {\bf 93} (2016) no.8,  084046;
  [arXiv:1602.06939 [gr-qc]].

\bibitem{Maselli:2016gxk}
  A.~Maselli, H.~O.~Silva, M.~Minamitsuji and E.~Berti,
  Phys.\ Rev.\ D {\bf 93} (2016) no.12,  124056;
  [arXiv:1603.04876 [gr-qc]].
  
\bibitem{Saito:2015fza}
  R.~Saito, D.~Yamauchi, S.~Mizuno, J.~Gleyzes and D.~Langlois,
  JCAP {\bf 1506} (2015) 008;
  [arXiv:1503.01448 [gr-qc]].

\bibitem{Babichev:2016jom}
  E.~Babichev, K.~Koyama, D.~Langlois, R.~Saito and J.~Sakstein;
  arXiv:1606.06627 [gr-qc].


\end{thebibliography}
\end{document}